\documentclass[12pt]{article}
\pdfoutput=1

\usepackage{putex}
\usepackage{graphicx}
\usepackage{caption}
\usepackage{amsmath}
\usepackage{array}
\usepackage{subcaption}
\usepackage{epstopdf}
\usepackage{enumerate}
\usepackage{cite}
\usepackage{dsfont}
\usepackage{youngtab}
\usepackage{tensor}
\usepackage{slashed}
\usepackage[aligntableaux=center]{ytableau}
\usepackage[utf8]{inputenc}
\usepackage{simplewick}
\usepackage[
      colorlinks=true,
      linkcolor=blue,
      urlcolor=blue,
      filecolor=black,
      citecolor=red,
      ]{hyperref}
\usepackage{braket}
\usepackage{mathtools}
\usepackage{rotating}
\usepackage{tabularx}
\newcolumntype{Y}{>{\centering\arraybackslash}X}
\newcolumntype{C}[1]{>{\centering\arraybackslash}p{#1}}
\usepackage{color, colortbl}
\definecolor{LightCyan}{rgb}{0.7,1,1}
\definecolor{Gray}{gray}{0.9}
\usepackage{tikz}
\usetikzlibrary{decorations.pathreplacing,decorations.markings,shapes.misc,arrows}
\usepackage{siunitx}

\newcommand{\grp}[1]{\mathrm{#1}}

\newcommand{\grU}{\grp{U}}
\newcommand{\grSU}{\grp{SU}}

\newcommand {\be} {\begin {equation}}
\newcommand {\ee} {\end {equation}}

\newcommand {\bes} {\begin {equation*}}
\newcommand {\ees} {\end {equation*}}

\newcommand{\es}[2] {\begin{equation} \label{#1} \begin{split} #2 \end{split} \end{equation}}

\newcommand{\Z}{\mathbb{Z}}

\newcommand{\beq}{\begin{equation}}
\newcommand{\eeq}{\end{equation}}

\def\ie{\begin{equation}\begin{aligned}}
\def\fe{\end{aligned}\end{equation}}

\numberwithin{equation}{section}

\def\<{\langle}
\def\>{\rangle}

\DeclareMathOperator{\SU}{SU}
\DeclareMathOperator{\SO}{SO}
\DeclareMathOperator{\U}{U}

\begin{document}

\preprint{PUPT-2656}

\institution{PU}{Joseph Henry Laboratories, Princeton University, Princeton, NJ 08544, USA}
\institution{PCTS}{Princeton Center for Theoretical Science, Princeton University, Princeton, NJ 08544, USA}
\institution{IAS}{School of Natural Sciences, Institute for Advanced Study, Princeton, NJ 08540, USA}

\title{Supercurrents and (Partial) Supersymmetry in Adjoint QCD$_2$ and Its Generalizations}

\authors{
Igor R.~Klebanov,\worksat{\PU, \PCTS} Silviu S.~Pufu,\worksat{\PU, \PCTS} Benjamin S\o gaard,\worksat{\PU} Edward Witten\worksat{\IAS}
}

\abstract{
$1+1$-dimensional $\SU(N)$ gauge theory coupled to an adjoint Majorana fermion, also known as adjoint QCD$_2$, has the surprising feature that at fermion mass $\sqrt{\frac{g^2 N}{2 \pi}}$ it exhibits supersymmetry.  In this paper, we obtain a deeper insight into how the supersymmetry works by constructing the gauge invariant, Lorentz covariant supercurrent $j_{\mu A}$. Its conservation relies crucially on the presence of a quantum anomaly.  We generalize this construction to a class of models where, in addition to an adjoint Majorana fermion of an appropriate mass, the gauge theory is coupled to some collection of massless fermions ($\SU(N)$ may be replaced by a more general gauge group). In general, these models have a supersymmetric massive sector and a non-supersymmetric CFT sector \cite{Popov:2022vud}, but there are cases in which both sectors are supersymmetric. An example of such a gapless, fully supersymmetric model is $\SU(N)$ gauge theory coupled to three adjoint Majorana fermions, of which two are massless and the third has 
mass  $\sqrt{\frac{3g^2 N}{2\pi}}$.
}

\maketitle

\tableofcontents

\section{Introduction and Summary}

The 't Hooft model \cite{tHooft:1974pnl} is a well-known $1+1$-dimensional toy model for Quantum Chromodynamics (QCD) that is solvable in the large $N$ limit. It is of further interest to replace the fermions in the fundamental representation by those in two-index representations; such $1+1$-dimensional models contain discrete analogues of $\theta$-vacua (flux tube sectors) \cite{Witten:1978ka}.  A minimal model of this type, the $\SU(N)$ gauge theory coupled to one adjoint Majorana fermion \cite{Dalley:1992yy}, is often called adjoint QCD$_2$. It has turned out to be an interesting playground for studying various non-perturbative phenomena in gauge theory. 
Its large $N$ spectrum consists of both bosonic and fermionic ``gluinoball" bound states that may be viewed as closed strings. Even though the light-cone bound state equations do not appear to be solvable analytically, some features of the spectra can be studied with good numerical precision   
\cite{Dalley:1992yy,Kutasov:1993gq,Bhanot:1993xp,Gross:1997mx,Katz:2013qua,Trittmann:2015oka,Dempsey:2021xpf}, even at finite $N$ \cite{Dempsey:2022uie}.  As we will review briefly, there are two values of the fermion mass $m$ at which this model exhibits enhanced symmetries and particularly interesting phenomena:  $m = 0$ and $m = \sqrt{\frac{g^2 N}{2 \pi}}$.

For generic values of $m$, the spectrum is gapped and the Wilson loop in the fundamental representation of $\SU(N)$ exhibits area law. As $m\to 0$, the mass gap remains, while, as was argued in \cite{Gross:1995bp}, the model makes a transition from the area law to perimeter law behavior of the fundamental Wilson loop.  Indeed, for $m=0$ the infrared physics is described by a gauged Wess-Zumino-Witten (WZW) model \cite{Witten:1983ar}, which is equivalent to a coset model \cite{Bardakci:1987ee}.  For a single Majorana fermion,
the coset model is $\frac{\SO(N^2-1)_1}{\SU(N)_{N}}$, and  
it has a vanishing Virasoro central charge \cite{Kutasov:1993gq}.  Thus, for any $N$, the $\SU(N)$ gauge theory coupled to a massless adjoint fermion serves as a non-trivial example of a gapped topological phase. 
An essential reason for the vanishing of string tension at $m=0$ is that the coset model 
$\frac{\SO(N^2-1)_1}{\SU(N)_{N}}$, which governs the long-distance behavior of adjoint QCD$_2$, is a topological quantum field theory and, therefore, does not have any notion of area. Therefore, the large Wilson loop cannot obey the area law.

The general properties of the $N$ flux tube sectors have been explored in \cite{Witten:1978ka,Smilga:1994hc,Lenz:1994du,Gross:1995bp,Smilga:1996dn,Cherman:2019hbq,Komargodski:2020mxz,Dempsey:2021xpf,Smilga:2021zrw}.  
As shown in \cite{Cherman:2019hbq,Komargodski:2020mxz}, for $m=0$ the number of vacua grows exponentially in $N$. These vacua, which are the states of the $\frac{\SO(N^2-1)_1}{\SU(N)_{N}}$ coset, are roughly evenly split between the $N$ flux tube sectors, and they are related by the action of certain non-invertible symmetries \cite{Komargodski:2020mxz}.  These nice observations have led to new arguments for the vanishing of the string tension at $m= 0$, with further evidence found in \cite{Dempsey:2021xpf,Dempsey:2023fvm,Dempsey:2024alw} using both light-cone and lattice gauge theory methods.

In this paper, we will be primarily interested in another remarkable property of adjoint QCD$_2$: 
for $m= \sqrt{\frac{g^2 N}{2\pi}}$, this theory becomes supersymmetric.\footnote{In some of the earlier literature, the supersymmetric value of mass was stated as $m= \sqrt{\frac{g^2 N}{\pi}}$ due to a different definition of the gauge coupling used there.} The emergence of supersymmetry is surprising, since the only dynamical degrees of freedom in this gauge theory are Majorana fermions. In this paper, we will obtain a deeper insight into how it works. The supersymmetry was discovered for large $N$ using light-cone quantization \cite{Kutasov:1993gq} (see also \cite{Bhanot:1993xp,Boorstein:1993nd,Antonuccio:1998zp,Dempsey:2021xpf}), and it was also found to hold at finite $N$ \cite{Dempsey:2022uie}.  The light-cone approach, however, has been limited to the trivial flux tube sector of the theory, and further checks were needed.  Recently, the supersymmetry was shown to be present on a small spatial circle with periodic boundary conditions \cite{Dempsey:2024ofo} in all the flux tube sectors. It is spontaneously broken in the non-trivial flux tube sectors, and numerical evidence for this breaking was found using Hamiltonian lattice gauge theory \cite{Dempsey:2023fvm,Dempsey:2024alw}.

More general gauge theories, where some massless fermion fields are present in addition to one massive adjoint fermion, were studied in \cite{Popov:2022vud}.  The physics of these models is typically described by a non-trivial coset Conformal Field Theory (CFT) and a massive sector. 
Using the non-Abelian bosonization \cite{Witten:1983ar} of the massless fermions, the supersymmetry of the level-$k$ gauged WZW model coupled to the massive adjoint fermion was demonstrated for
$m= \sqrt{\frac{g^2 (N+k)}{2\pi}}$.\footnote{If the gauge coupling is set to zero, this reduces to the supersymmetry of the WZW model coupled to a massless adjoint fermion \cite{DiVecchia:1984nyg}.}
 However, typically only the massive sector of the full theory is supersymmetric, while the CFT sector is not \cite{Popov:2022vud}. Therefore, the generalized models of 
\cite{Popov:2022vud} can be called ``partially supersymmetric."

In this paper, we derive the supersymmetry of adjoint QCD$_2$ in a gauge invariant way by presenting an explicit, Lorentz covariant expression for the supercurrent $j_{\mu A}$ and checking its conservation when $m= \sqrt{\frac{g^2 N}{2\pi}}$.  The supercurrent conservation relies in a crucial way on the quantum anomalies. An important feature of the supercurrent is that, in addition to the term cubic in the adjoint fermion, it contains a term where the field strength $F$ is multiplied by the fermion. In the non-trivial flux tube sectors, where $F$ has a non-zero background value, the resulting term linear in the fermion gives rise to spontaneous breaking of supersymmetry \cite{Witten:1982df}. 

We also construct the conserved supercurrent for the generalized models of  \cite{Popov:2022vud}
and show that it is conserved for $m= \sqrt{\frac{g^2 (N+k)}{2\pi}}$. In these cases, the energy-momentum tensor $T^\text{SUSY}_{\mu \nu}$ that belongs to the same supermultiplet as the supercurrent generally contains certain four-fermion terms and differs from the canonical energy-momentum tensor by the stress tensor of the coset CFT\@. This is the reason why only the massive spectrum is supersymmetric in these models.  However, if the massless fermions are chosen in such a way that the infrared (IR) Virasoro central charge, $c_{\rm IR}$, vanishes and the CFT sector becomes topological (a full list of such models was presented in \cite{Delmastro:2021otj}), then the model becomes fully supersymmetric.   One simple example of such a fully supersymmetric model, discussed in Section~\ref{TwoAdjoint}, is $\SU(N)$ gauge theory
coupled to two adjoint fermions, one massless and the other of mass  $\sqrt{\frac{g^2 N}{\pi}}$. It is not hard to check the cancellation of the 4-fermion terms 
in this model.

For other special choices of massless fermionic matter where the $c_{\rm IR}>0$ coset CFT is supersymmetric, the gauge theory may be again fully supersymmetric. In Section~\ref{ThreeAdjoint} we present such an ${\cal N}=(1, 1)$ supersymmetric {\it gapless} $\SU(N)$ gauge theory. It contains three adjoint fermions, two of them massless and the third of 
mass  $\sqrt{\frac{3g^2 N}{2\pi}}$. The low-energy limit of this theory is described by the coset CFT $\frac{\SO(2N^2-2)_1}{\SU(N)_{2N}}$, which has ${\cal N}=(2, 2)$ supersymmetry \cite{Gopakumar:2012gd,Isachenkov:2014zua}; its non-invertible symmetries were analyzed in \cite{Damia:2024kyt}.

The rest of this paper is organized as follows.  In Section~\ref{ADJOINT} we study the adjoint QCD$_2$ model, prove the conservation of the supercurrent, and explore the supercurrent multiplet.  In Section~\ref{MASSLESS} we introduce additional massless fermions and explore the family of partially supersymmetric theories. We also generalize our construction of the supercurrent to other gauge groups. For the $\U(1)$ gauge group, discussed in Section~\ref{Abelian}, conservation of the supercurrent follows from the chiral anomaly equation in the Schwinger model \cite{Schwinger:1962tp}. 
Several technical details are relegated to the Appendices.

\section{Supersymmetry in adjoint QCD$_2$}
\label{ADJOINT}

\subsection{Action and classical equations of motion}

When working with fermions in $1+1$ dimensions, we will use the gamma matrices $\gamma^0 = \sigma_2$, $\gamma^1 = i\sigma_1$, obeying the Clifford algebra $\{ \gamma^\mu, \gamma^\nu\} = 2 \eta^{\mu\nu}$, with $\eta^{\mu\nu} = \diag\{ 1, -1\}$.  With these choices, the chirality matrix is $\gamma^5 = \gamma^0 \gamma^1 = \sigma_3$, and the components of Majorana spinors are real. The action for our model for an $\SU(N)$ gauge theory with gauge field $A_\mu$ and an adjoint Majorana fermion $\psi$ is 
\es{STotal}{
    S  = \int d^2x \tr\left(-\frac{1}{2 g^2}F_{\mu\nu}F^{\mu\nu} + i
\overline{\psi}\slashed{D}\psi - m  \overline{\psi}\psi\right) \, ,
} 
 where $g > 0$ is the gauge coupling, $m$ is the mass, the trace is taken in the fundamental representation of $\SU(N)$, and the gauge covariant derivative acting on any adjoint-valued field $X$ is defined by 
 \es{CovDer}{
    D_\mu X = \partial_\mu X - i [A_\mu, X]   \,.
 }
For an adjoint-valued field $X$, we can define the color components $X^a$ via $X = X^a T^a$, with $T^a$ being the generators of $\SU(N)$ normalized such that $\tr (T^a T^b) = \frac{\delta^{ab}}{2}$ in the fundamental representation.   The axial $\Z_2$ transformation $\psi \to \gamma^5 \psi$ can be used to send $m \to -m$.\footnote{There are 't Hooft anomalies between this transformation and the center symmetry, the charge conjugation symmetry, and the fermion parity \cite{Cherman:2019hbq}.  Thus, while the full energy spectrum of the theory is the same at mass $m$ and at mass $-m$, the labeling of the flux tube sectors, of the charge conjugation assignments, and of which states are bosons and which are fermions may change when we flip the sign of $m$.}  We therefore assume $m \geq 0$ without loss of generality in the rest of this paper.

We define the light-cone coordinates $x^\pm = (x^0 \pm x^1) / \sqrt{2}$ as well as the light-cone components of the gauge field $A_\pm = (A_0 \pm A_1) / \sqrt{2}$.  The spacetime indices can be raised and lowered with the metric, whose only non-vanishing components are $\eta_{+-}= \eta_{-+}=1$.  For the fermions, we use the chiral components $\psi_\pm$ defined by
 \es{PsiDef}{
  \psi = \frac{1}{2^{1/4}}\begin{pmatrix} \psi_{-} \\ \psi_{+}  \end{pmatrix} \,.
 }
In $1+1$ dimensions, the electric field is a Lorentz scalar, which is made explicit by defining $F \equiv -\frac 12 \epsilon^{\mu\nu} F_{\mu\nu}$, with $\epsilon^{01} = - \epsilon_{01} = 1$.  In light-cone components, we have $F = F_{+-}$.

With the notation established above, the action \eqref{STotal} becomes 
 \es{Action}{
   S = \int d^2x\, \tr\left(\frac{1}{g^2} F^2 +  i\psi_- \partial_+ \psi_- +  i\psi_+ \partial_- \psi_+ - 2 A_+ J_- - 2 A_- J_+ - i \sqrt{2} m \psi_+ \psi_-\right)  \,, 
 }
where the gauged $\SU(N)$ current $J_\mu  =\overline{\psi} \gamma_\mu \psi$ is written in light-cone components as\footnote{Note that our sign convention for $J_\mu$ is opposite to some of the literature, in particular \cite{Peskin:1995ev}.}
 \es{JDefs}{
 J_- \equiv  \psi_- \psi_-  \,, \qquad  J_+ \equiv  \psi_+ \psi_+  \,.
}
In these formulas all quantities are Lie-algebra valued, so for instance the equation $J_- = \psi_- \psi_-$ is understood as $J_-^a T^a = \psi_-^b \psi_-^c T^b T^c$.  Due to the fact that $\psi_-^b$ are Grassmann-valued fields, we can also write this expression as $\frac 12 \psi_-^b \psi_-^c [T^b, T^c] = \frac i2 f^{abc} \psi_-^b \psi_-^c T^a$.  Thus, the color components of $J_-$ are $J_-^a = \frac{i}{2} f^{abc} \psi_-^b \psi_-^c$,\footnote{Equivalently, one can think of $\psi_\pm$ and $J_\pm$ as $N\times N$ matrix-valued fields, with matrix multiplication being understood in \eqref{JDefs}.}  and, similarly, $J_+^a = \frac{i}{2} f^{abc} \psi_+^b \psi_+^c$.  
The equations of motion that follow from the action \eqref{Action} are
 \es{eoms}{
    J_- = \frac{1}{g^2}D_- F \,, \qquad  J_+ = -\frac{1}{g^2} D_+ F \,, \qquad
     D_-\psi_+ = \frac{m}{\sqrt{2}} \psi_- \,, \qquad   D_+\psi_- = -\frac{m}{\sqrt{2}} \psi_+ \,.
 }

These equations and many of the ones that follow are constrained by Lorentz symmetry and parity, which are symmetries of the adjoint QCD$_2$ theory.  Boosts send
 \es{APsiLorentz}{
  A_-(x^+, x^-) &\to \lambda A_-(\lambda^{-1} x^+, \lambda x^-) \,, \qquad \ \ \ 
    A_+(x^+, x^-) \to \lambda^{-1} A_+(\lambda^{-1} x^+, \lambda x^-) \,, \\
  \psi_-(x^+, x^-) &\to \lambda^{1/2} \psi_-(\lambda^{-1} x^+, \lambda x^-) \,, \qquad 
    \psi_+(x^+, x^-) \to \lambda^{-1/2} \psi_+(\lambda^{-1} x^+, \lambda x^-) \,,
 }
with $\lambda$ an arbitrary parameter.  All other fields with spin transform similarly based on their index structure.  Lorentz covariance implies that the difference between the number of lower $-$ spacetime indices plus half the number of left-chiral ($-$) spinor components and the number of lower $+$ spacetime indices plus half the number of right-chiral ($+$) spinor components must be the same on both sides of each equation. One can easily check that the equations of motion in \eqref{eoms} obey these constraints. Parity swaps the $+$ and $-$ light-cone indices and acts as $i \gamma^0= i \sigma_2$ on spinors.  In particular, 
 \es{APsiParity}{
  A_- (x^+, x^-) &\to A_+(x^-, x^+) \,, \qquad  A_+ (x^+, x^-) \to A_-(x^-, x^+) \,, \\
  \psi_-(x^+, x^-) &\to \psi_+(x^-, x^+) \,, \qquad \, \psi_+(x^+, x^-) \to -\psi_-(x^-, x^+) \,.
 }
The field strength $F$ is a pseudo-scalar field, since $F(x^+, x^-) = F_{+-} (x^+, x^-) \to F_{-+}(x^-, x^+) = - F(x^-, x^+)$.  One can easily check that the second and third equations of motion in \eqref{eoms} are obtained, respectively, from the first and fourth.

Standard formulas for the stress tensor give, in light-cone components,
 \es{TComponents}{
  T_{--} &= i \tr ( \psi_- D_- \psi_-) \,, \\
  T_{++} &= i \tr (\psi_+ D_+ \psi_+) \,, \\
  T_{+-}  &=  \frac{1}{g^2} \tr F^2 +   \frac{i m}{\sqrt{2}} \tr (\psi_+ \psi_-) = T_{-+} \,.
 }
The stress tensor is conserved upon using the equations of motion \eqref{eoms}.\footnote{For instance, $\partial_+ T_{--} =  -\frac{im}{\sqrt{2}}  \tr ( \psi_+ D_- \psi_-)  + i \tr ( \psi_- [D_+, D_-] \psi_-)$ and $\partial_- T_{+-} =   2 \tr (F  \psi_- \psi_-) + \frac{i m}{\sqrt{2}} \tr (\psi_+ D_- \psi_-)$, up to terms that trivially vanish using the trace identities.  Since $[D_+, D_-] \psi_- = -i [F, \psi_-]$, we have that $\partial_+ T_{--} + \partial_- T_{+-} = 0$.  The other conservation equation, $\partial_- T_{++} + \partial_+ T_{-+} = 0$, follows from the previous one by parity symmetry.
 }  As usual, the momentum operator $P_\mu$ can be obtained by integrating $T^0{}_\mu$ over a spatial slice:
 \es{PEqualTime}{
  P_\mu = \int dx\, T^0{}_\mu = \int dx\, \frac{T_{+\mu} + T_{-\mu}}{\sqrt{2}}  \,.
 }
Thus, from \eqref{TComponents}, using the equations of motion we arrive at the expressions
 \es{GotPpm}{
  P_+ &= \int dx \,  \left[   i \tr (\psi_+ D_1 \psi_+)  + \frac{1}{\sqrt{2} g^2} \tr F^2 +  im \tr (\psi_+ \psi_-) \right]  \,, \\
  P_- &= \int dx \,  \left[  -  i \tr (\psi_- D_1 \psi_-) + \frac{1}{\sqrt{2} g^2} \tr F^2 + i m  \tr (\psi_+ \psi_-) \right]  \,.
 }

\subsection{Quantum anomalies for composite operators} 

The discussion so far has been at the classical level, but \eqref{eoms} and the equations that follow also hold in the quantum theory as operator equations after appropriate regularization.  In the quantum theory, extra care must be taken when defining composite operators such as $J_\mu$ or $T_{\mu\nu}$.  In general, when taking further derivatives of composite operators, we may encounter extra contributions due to quantum anomalies that cannot be seen classically.

For the discussion that follows, it will be sufficient to consider composites of two adjoint-valued operators ${\cal A} = {\cal A}^a T^a$ and ${\cal B} = {\cal B}^a T^a$.  We will consider a singlet composite ${\cal X}$ and an adjoint composite ${\cal Y} = {\cal Y}^a T^a$ defined schematically by
 \es{XYDefs}{
  {\cal X} \equiv \tr ({\cal A} {\cal B}) = \frac 12 {\cal A}^a {\cal B}^a \,, \qquad
   {\cal Y}^a \equiv i f^{abc} {\cal A}^b {\cal B}^c \,.
 }
For the case where ${\cal A}$ and ${\cal B}$ do not have short-distance singularities with $A_\mu$, more precise definitions can be made by regularizing the composite operators by point-splitting their constituents and inserting a Wilson line in the adjoint representation between them in order to preserve gauge invariance (for ${\cal X}$) or gauge covariance (for ${\cal Y}$):
 \es{ABReg}{
  {\cal X}(x) &= \frac 12 \lim_{\epsilon \to 0}  \left( {\cal A}^a(x) U^{ab}(x, x+\epsilon)  {\cal B}^b(x+\epsilon)  \right) \,, \\
 {\cal Y}^a(x) &= i f^{abc} \lim_{\epsilon \to 0} \left( {\cal A}^b(x) U^{cd} (x, x+\epsilon) {\cal B}^d(x+\epsilon) \right) \,.
 }
Here, $\epsilon^\mu$ is a space-time vector corresponding to the point splitting, and 
 \es{UDef}{
  U(x_1, x_2) \equiv P \exp \left[ -i \int_{x_1}^{x_2} dz^\mu\, A_\mu(z) \right]
 }
is the parallel transport operator from $x_2$ to $x_1$, which can in principle be evaluated in any representation, with $U^{ab}(x_1, x_2)$ being its matrix elements in the adjoint representation.

A regularization prescription that breaks Lorentz invariance is to keep the direction of $\epsilon^\mu$ fixed as we take $\epsilon^\mu \to 0$.  For instance, if we perform equal-time quantization, it is convenient to split the points in the space direction, so we consider $\epsilon^\mu = (0, \epsilon^1)$ and take $\epsilon^1 \to 0$.  Such a regularization prescription arises naturally when composite operators appear in equal-time commutators or anti-commutators.  Alternatively, a Lorentz-invariant regularization prescription can be obtained by averaging over the direction of $\epsilon^\mu$ using (see Chapter 19.1 of \cite{Peskin:1995ev})
 \es{LIAvg}{
  \text{Lorentz-invariant prescription:} \qquad
   \lim_{\epsilon \to 0} \frac{\epsilon^\mu}{\epsilon^2} = 0 \,, \qquad
    \lim_{\epsilon \to 0} \frac{\epsilon^\mu \epsilon^\nu}{\epsilon^2} = \frac{\eta^{\mu\nu}}{2} \,, \qquad \text{etc.} 
 }
Unless otherwise specified, we will use the regularization where point-splitting is performed in the space direction.  We can interpret the operators in \eqref{JDefs} and \eqref{TComponents} in this manner.

The derivatives of the composite operators with respect to $x^\mu$ will have the usual terms expected by the product rule, as well as extra terms arising from taking derivatives of the gauge field in the Wilson line.  These derivatives yield finite contributions provided that the product ${\cal A}(x) {\cal B}(x+ \epsilon)$ has a linearly-divergent contribution in $\epsilon$, which can be extracted from the OPE\@.

We will encounter two cases.  The first case is for the adjoint composite ${\cal Y}$ when the OPE of the operators ${\cal A}$ and ${\cal B}$ takes the form
 \es{OPE1}{
  {\cal A}^a(x) {\cal B}^b(0) =  \frac{1}{x^-} \delta^{ab} {\cal C}(0) + \cdots 
   \qquad 
   \text{or}  
    \qquad
  {\cal A}^a(x) {\cal B}^b(0) = - \frac{1}{x^+} \delta^{ab} {\cal C}(0) + \cdots  \,, 
 } 
for some gauge-invariant operator ${\cal C}$, where the ellipses denote less singular OPE contributions.  Note that the first terms on the right-hand sides of \eqref{OPE1} seem to break gauge covariance, because these terms do not transform under gauge transformations in the same way as the left-hand sides of the corresponding equations.  Of course, in each case gauge covariance is restored by the full OPE\@.  In particular, the leading term in the first expression, $\frac{1}{x^-} \delta^{ab} {\cal C}(0)$, should be interpreted as a leading approximation to $\frac{1}{x^-} U^{ab}(x, 0) {\cal C}(0)$, where $U^{ab}(x, 0)$ is the parallel transport operator \eqref{UDef} (a Wilson line) in the adjoint representation.  Unlike $\frac{1}{x^-} \delta^{ab} {\cal C}(0)$, the expression $\frac{1}{x^-} U^{ab}(x, 0) {\cal C}(0)$ does have the same gauge transformation properties as the left-hand side of the corresponding equation.\footnote{A similar comment applies to all OPEs used in this work, namely the terms we write down should be understood as being completed into an expression with the right gauge transformation properties once the higher-order terms in the OPEs are included.}  Nevertheless, with the OPEs as in \eqref{OPE1}, we show in Appendix~\ref{ANOMALYAPPENDIX} that
 \es{DerAB1}{
  (D_+ {\cal Y})^a &= i f^{abc} (D_+ {\cal A})^b {\cal B}^c + i f^{abc} {\cal A}^b (D_+ {\cal B})^c - i  N  F^a {\cal C} \\  
       &\text{or}  \\
  (D_- {\cal Y})^a &= i f^{abc} (D_- {\cal A})^b {\cal B}^c + i f^{abc} {\cal A}^b (D_- {\cal B})^c - i  N  F^a {\cal C} \,,
 }
where the two expressions correspond, respectively, to the two cases in \eqref{OPE1}.  These expressions hold both for the non-Lorentz-invariant regularization prescription in which we keep the direction of $\epsilon^\mu$ fixed, as well as for the prescription that includes the Lorentz-averaging \eqref{LIAvg}.\footnote{The extra contribution to $(D_- {\cal Y})^a$ in the first case of \eqref{OPE1} is proportional to $\epsilon^+/\epsilon^- = 2 (\epsilon^+)^2 / \epsilon^2$.  It does not vanish if we use a Lorentz-non-invariant regulator (in particular, for spatial point-splitting), but it vanishes upon the averaging \eqref{LIAvg}.  However, such a derivative will not be needed in the discussion below.   Similar comments holds for $(D_+ {\cal Y})^a$ in the second case of \eqref{OPE1}.\label{NonLFootnote}}

The second case we will be interested in, which will be relevant in the next subsection, is for the derivatives of the singlet composite ${\cal X}$ when the OPE of the operators ${\cal A}$ and ${\cal B}$ takes the form
 \es{OPE2}{
  {\cal A}^a(x) {\cal B}^b(0) = \frac{1}{x^-} f^{abc} {\cal C}^c(0) + \cdots  
   \qquad \text{or} \qquad
  {\cal A}^a(x) {\cal B}^b(0) = - \frac{1}{x^+} f^{abc} {\cal C}^c(0) + \cdots \,.
 }
Similarly to the previous case, we have 
 \es{DerAB2}{
  \partial_+ {\cal X} &= \tr \left( (D_+ {\cal A}) {\cal B} \right) 
   + \tr \left( {\cal A} (D_+ {\cal B}) \right)  + N \tr \left( F {\cal C}  \right)  \\
   &\text{or} \\
   \partial_- {\cal X} &= \tr \left( (D_- {\cal A}) {\cal B} \right) 
   + \tr \left( {\cal A} (D_- {\cal B}) \right)  + N \tr \left( F {\cal C}  \right)  \,,
 }
where the two expressions again correspond to the two cases in \eqref{OPE2}.  The relations \eqref{DerAB2} also hold for both the non-Lorentz-invariant and Lorentz-averaged regularization prescriptions, as we show in Appendix~\ref{ANOMALYAPPENDIX}.\footnote{A similar comment to that in Footnote~\ref{NonLFootnote} applies:  the extra contribution to $\partial_- {\cal X}$ in the first case of \eqref{OPE2} is proportional to $\epsilon^+/\epsilon^- = 2 (\epsilon^+)^2 / \epsilon^2$.  It does not vanish for spatial point-splitting, but it vanishes upon Lorentz-averaging.  A similar comment holds for $\partial_+ {\cal X}$ in the second case of \eqref{OPE2}.}

An immediate application of \eqref{DerAB1} is to the computation of $D_+ J_-$ and $D_- J_+$, which are composites of two fermions.  The OPE of the two fermion fields is 
 \es{OPEFerm}{
  \psi^a(x) \overline{\psi}^b(0) = - \frac{i \delta^{ab} }{2 \pi} \frac{\gamma^\mu x_\mu}{x^2} + \cdots \,,
 }
which implies  $\psi^a_-(x) \psi^b_-(0) = -\frac{i \delta^{ab}}{2 \pi x^-} + \cdots$ and $\psi^a_+(x) \psi^b_+(0) = -\frac{i \delta^{ab}}{2 \pi x^+} + \cdots$.  As per \eqref{JDefs}, we have $J_-^a = \frac i2  f^{abc} \psi_-^b \psi_-^c$, so we can apply \eqref{DerAB1} with ${\cal A} = {\cal B} = \psi_-$ and ${\cal C} = -\frac{i}{2 \pi}$.  Also using the equation of motion $D_+ \psi_- = - \frac{m}{\sqrt{2}} \psi_+$, we find
 \es{DJp}{
  D_+ J_- = - \frac{m}{\sqrt{2}} (\psi_+ \psi_- + \psi_- \psi_+) - \frac{N}{4 \pi} F \,.
 }
Similarly, from $J_+^a = \frac i2 f^{abc} \psi_+^b \psi_+^c$ and the equation of motion $D_- \psi_+ = \frac{m}{\sqrt{2}} \psi_-$, we can use \eqref{DerAB1} with ${\cal A} = {\cal B} = \psi_+$ and ${\cal C} = \frac{i}{2 \pi}$ to find
 \es{DJm}{
  D_- J_+ =  \frac{m}{\sqrt{2}} (\psi_+ \psi_- + \psi_- \psi_+) + \frac{N}{4 \pi} F \,.
 }
(The same expression can be obtained from \eqref{DJp} using a parity transformation.)  The equations \eqref{DJp} and \eqref{DJm} are consistent with the fact that the gauged current is conserved:  $D^\mu J_\mu = D_+ J_- + D_- J_+ = 0$.  The $\SU(N)$ axial current $J_\mu{}^5 = \overline{\psi} \gamma_\mu \gamma^5 \psi = - \epsilon_{\mu\nu} J^\nu$, on the other hand, is not conserved even for massless fermions, since $D^\mu J_\mu{}^5 = -D_- J_+ + D_+ J_- = -\sqrt{2} m (\psi_+ \psi_- + \psi_- \psi_+) - \frac{N}{2\pi} F$.

\subsection{Supercurrent}

As mentioned in the Introduction, when $m = \sqrt{\frac{g^2 N}{2 \pi}}$, it was shown using light-cone quantization that the adjoint QCD$_2$ theory exhibits ${\cal N} = (1, 1)$ supersymmetry \cite{Kutasov:1993gq,Boorstein:1993nd}.  In the light-cone approach one usually works in the gauge $A_- = 0$ and integrates out $A_+$ and $\psi_+$, so the supersymmetry is not exhibited in a spacetime covariant way in an arbitrary gauge.  In this section, we will write down a gauge-invariant local ansatz for the corresponding supercurrent $j_{\mu A}$, where 
$A$ is the spinor index.  We impose the conservation condition for it and we rederive the value of $m$ at the supersymmetric point mentioned above.  

Like for the fermion field $\psi_A$, for the supercurrent $j_{\mu A}$ we use the chiral spinor components:
\es{jChiral}{
 j_\mu = \frac{1}{2^{1/4}} \begin{pmatrix}
  j_{\mu-} \\
  j_{\mu+}
\end{pmatrix} \,.
}
In light-cone coordinates, we start with the ansatz
 \es{LCsupercurrents}{
  j_{--} &= \frac{1}{3} \tr \psi_-^3 = \frac 13 \tr (\psi_- J_-) \,,  \qquad j_{+-} = \alpha \tr \left( \psi_+ F \right) \,, \\
  j_{++} &= \frac{1}{3} \tr \psi_+^3 = \frac 13 \tr (\psi_+ J_+) \,, \qquad  j_{-+} = \alpha \tr \left( \psi_- F \right) \,,
 }
for some constant $\alpha$ that we will need to determine.   In a more covariant notation, this ansatz can be written as
  \es{SuperCurrentTot}{
  j_\mu =\frac 16 \tr \left( 
J_\nu \gamma^\nu \gamma_\mu \psi  \right) 
+ i \frac{\alpha }{\sqrt{2}}  \tr \left(F_{\mu\nu} \gamma^\nu \psi \right) \,.
 }
As we will now show, the supercurrent is conserved when we take 
 \es{normalizations}{
  m =  \sqrt{\frac{g^2 N}{2\pi}} \,,
    \qquad 
    \alpha = \sqrt{\frac{N} {4 g^2 \pi}} \,,
 }
thus proving that the theory possesses ${\cal N} = (1, 1)$ supersymmetry at this value of the mass.

Let us see how \eqref{normalizations} arises from the conservation equation for one of the chirality components of the supercurrent,
 \es{pluscomp}{
  \partial_+ j_{--} + \partial_- j_{+-}=0 \,.
 }
As we saw in the previous subsection, we should be careful when taking derivatives of composite operators due to anomaly terms that might appear.  Since the $F \times \psi_+$ OPE does not have any singular contributions, for the second term in \eqref{pluscomp} we can just use the product rule and the equations of motion \eqref{eoms}: 
 \es{djSecond}{
\partial_- j_{+-} = \alpha \tr ((D_-\psi_+) F) + \alpha \tr \left( \psi_+ D_-F \right) = \alpha \frac{m}{\sqrt{2}} \tr \left( \psi_-   F\right)  + \alpha g^2 \tr \left( \psi_+ \psi_- \psi_- \right) \,.
 }
For the first term in \eqref{pluscomp}, we can use \eqref{DerAB2} with ${\cal A}^a = \psi_-^a$ and ${\cal B}^a = J_-^a = \frac{i}{2} f^{abc} \psi_-^b \psi_-^c$.  Since the singular term in the $\psi_-^a \times \psi_-^b$ OPE is $-\frac{i}{2 \pi x^-} \delta^{ab}$, the singular term in the $\psi_- \times J_-$ OPE is 
 \es{psiJmOPE}{
  \psi_-^a(x) J_-^b(0) = - f^{abc} \frac{1}{2 \pi x^-} \psi_-^c(0) + \cdots \,,
 }  
and so we identify ${\cal C} = - \frac{1}{2 \pi} \psi_-$.  The equation \eqref{DerAB2} then reads
  \es{DerAB2Applied}{
  \partial_+ \tr (\psi_- J_-) = \tr \left( (D_+ \psi_-) J_- \right) 
   + \tr \left( \psi_- (D_+ J_-) \right)  - \frac{N}{2 \pi} \tr \left( F \psi_-  \right)  \,.
 }
Dividing by $3$ and using \eqref{eoms} and \eqref{DJp}, we find 
 \es{djQuantum}{
  \partial_+ j_{--} = - \frac{m}{\sqrt{2}} \tr \left( \psi_+ \psi_- \psi_- \right)  - \frac{N}{4 \pi} \tr \left( \psi_- F \right) \,.
 }
This expression precisely cancels the one in \eqref{djSecond} when $\alpha \frac{m}{\sqrt{2}} = \frac{N}{4 \pi}$ and $\alpha g^2 =  \frac{m}{\sqrt{2}}$.  The solution of these equations for $m$ and $\alpha$ that obeys $m \geq 0$ is given by \eqref{normalizations}.  The other component of the supercurrent conservation equation,
 \es{minuscomp}{
  \partial_+ j_{-+} + \partial_- j_{++}=0 \,,
 }
can be checked analogously. 

The supercharge can be obtained by integrating the time component of the supercurrent over a spatial slice:
 \es{QEqualTime}{
  Q = \int dx\, j^0 = \int dx\, \frac{j_+ + j_-}{\sqrt{2}} 
   =  \frac{1}{2^{1/4}} \begin{pmatrix} Q_- \\ Q_+ \end{pmatrix} \,,
 }
where
 \es{GotQEqualTime}{
  Q_- &= \int dx\, \frac{1}{\sqrt{2}} \left[ \frac{1}3 \tr \left( \psi_-^3  \right)  +  \alpha \tr (F \psi_+)  \right] \,, \\
  Q_+ &= \int dx\, \frac{1}{\sqrt{2}}  \left[   \frac {1}3 \tr \left(\psi_+^3 \right) + \alpha \tr (F \psi_- ) \right]  \,.
 }
In light-cone quantization, where $x^-$ is interpreted as space and $x^+$ as time, the expressions for the supercharges take the simpler forms $Q_-^{\rm LC} =  \int dx^-\, j_{--} = \int dx^- \, \frac 13 \tr \psi_-^3$ and $Q_+^{\rm LC} = \int dx^- \, j_{-+} = \alpha \int dx^- \, \tr (\psi_- F)$.  These expressions agree with those found in \cite{Kutasov:1993gq,Boorstein:1993nd} in the light-cone approach obtained in the gauge $A_- = 0$ after integrating out $A_+$ and $\psi_+$.

The presence of terms containing the field strength $F$ in the supercharge (\ref{QEqualTime}) has an important physical consequence. In the non-trivial flux tube sectors, where $F$ has a non-zero background value, the resulting term linear in the fermion gives rise to spontaneous breaking of supersymmetry \cite{Witten:1982df}. This is expected on general grounds due to the non-zero string tension at the supersymmetric point \cite{Dubovsky:2018dlk} and was observed explicitly in the small circle \cite{Dempsey:2024ofo} and Hamiltonian lattice calculations \cite{Dempsey:2023fvm,Dempsey:2024alw}.

\subsection{The supercurrent multiplet}
\label{MULTIPLET}

As in any supersymmetric theory, the supercurrent belongs to a supercurrent multiplet that also contains the stress-energy tensor.  The 2D supercurrent multiplets were discussed in \cite{Dumitrescu:2011iu,Chang:2018dge}.  The most general $(1,1)$ multiplet contains the stress tensor $T_{\mu\nu}$, the supercurrent $j_{\mu A}$, and a conserved current $Z_\mu$ associated with a scalar central charge.  The supersymmetry transformations, implemented at the quantum level by (anti-)commutators with the supercharge, take the form
 \es{SUSYTransf}{
  [Q, Z_\mu] = c_1 \epsilon_{\mu\rho} \partial^\rho \gamma^\nu j_\nu \,, \qquad
   \{ j_\mu, \overline{Q} \} = c_2 T_{\mu\nu} \gamma^\nu + c_3 \gamma^5 Z_\mu \,,
 }
for constants $c_1$, $c_2$, and $c_3$.  As we will see, the supermultiplet that arises in 2D adjoint QCD$_2$ is of a special type where the current associated with the central charge is the dual of a scalar operator $Z$, namely $Z_\mu = \epsilon_{\mu\rho} \partial^\rho Z$.   This expression implies that the central charge of the ${\cal N}=(1, 1)$ supersymmetry
algebra is
 \es{calZDef}{
  {\cal Z}=  Z(x^0, \infty) - Z(x^0, -\infty)  \,.
 }
The central charge can be non-vanishing only for domain wall solutions that interpolate between vacua that have different $\langle Z \rangle$, if any such domain walls exist.  The supersymmetry transformations are
 \es{SUSYTransf2}{
  [Q, Z ] = c_1 \gamma^\nu j_\nu \,, \qquad
   \{ j_\mu, \overline{Q} \} = c_2 T_{\mu\nu} \gamma^\nu + c_3 \gamma^5 \epsilon_{\mu\rho} \partial^\rho Z \,.
 } 

As we will now show, the scalar operator in the supercurrent multiplet is
 \es{ZDef}{
  Z = \frac{\alpha}{2 g^2} \tr (F^2) \,,
 }
and the constants $c_i$ in \eqref{SUSYTransf2} are given by
 \es{Gotcs}{
  c_1 = \frac{N}{8 \sqrt{2} \pi} \,, \qquad
   c_2 = \frac{N}{16 \pi} \,, \qquad c_3 = - \frac{i}{2 \sqrt{2}} \,.
 }
In light-cone coordinates, the relations \eqref{SUSYTransf2} with the constants \eqref{Gotcs} can be written as
 \es{QZExplicit}{
  [Q_-, Z ] =  -\frac{i N}{8 \pi } j_{-+} \,, \qquad
   [Q_+, Z ]  = \frac{i N}{8 \pi } j_{+-}  \,,
 }
for the first equation in \eqref{SUSYTransf2}, and
 \es{QjExplicit}{
     \{ Q_-, j_{--}\} &=  \frac{N}{8 \pi} T_{--} \,, \qquad
  \{ Q_-, j_{-+}\} =  \frac{1}{2} \partial_- Z \,, \\
  \{ Q_-, j_{+-} \} &=  
    \frac{N}{8 \pi}  T_{+-} \,, \qquad
  \{ Q_-, j_{++} \} =  -\frac{1}{2} \partial_+ Z \,, \\
   \{Q_+, j_{++} \} &=  \frac{N}{8 \pi} T_{++} \,, \qquad
  \{ Q_+, j_{+-}\} =  -\frac{1}{2} \partial_+ Z \,, \\ 
  \{ Q_+, j_{-+} \} &=  
    \frac{N}{8 \pi}  T_{+-} \,, \qquad
   \{ Q_+, j_{--} \} =  \frac{1}{2} \partial_- Z \,,
 }
for the second equation in \eqref{SUSYTransf2}.

The commutation relations \eqref{QZExplicit}--\eqref{QjExplicit} can be derived in equal-time quantization using the anti-commutation relations
 \es{CanRels}{
  \{ \psi_-^a(t, x), \psi_-^b(t, y) \} =  \{ \psi_+^a(t, x), \psi_+^b(t, y) \} = \sqrt{2} \delta^{ab} \delta(x-y)  \,,
 }
as well as the additional commutation relations 
  \es{FAdditional}{
  \left[F^a(t, x), \frac 13 \tr (\psi_-^3)(t, y)\right] &=   \frac{ig^2 N}{4\sqrt{2}\pi}   \delta(x-y)  \psi_-^a(t, y) \,, \\
  \left[F^a(t, x), \frac 13 \tr (\psi_+^3)(t, y)\right] &=   -\frac{ig^2 N}{4\sqrt{2}\pi}   \delta(x-y)  \psi_+^a(t, y) 
 }
that come from the fact that the operators $\tr (\psi_-^3) = \tr (\psi_- J_-)$ and $\tr (\psi_+^3) = \tr (\psi_+ J_+)$ are regularized by point splitting with an additional Wilson line between the two points that does not commute with $F^a$.  We derive these expressions in Appendix~\ref{ANOMALYAPPENDIX}.  We list all the relevant commutators and anti-commutators required for the supersymmetry algebra in \eqref{RelevantComm} and \eqref{RelevantAnticomm} in Appendix~\ref{COMMUTATORS}.  Using the expressions for the supercharges \eqref{GotQEqualTime}, the supercurrent \eqref{LCsupercurrents}, the stress tensor \eqref{TComponents}, the definition \eqref{ZDef} of the operator $Z$, and the equations of motion \eqref{eoms}, it is straightforward to reproduce \eqref{QZExplicit} and \eqref{QjExplicit}.  A few more intermediate steps are given in Appendix~\ref{COMMUTATORS}.

From \eqref{QjExplicit} we can determine the anti-commutators of the supercharges by integrating the time component of the supercurrent over $x$.  We obtain
 \es{SUSYAlg}{
  \{Q_-, Q_- \} = \frac{N}{8 \pi} P_- \,, \qquad \{Q_+, Q_+ \} = \frac{N}{8 \pi} P_+ \,, \qquad
   \{Q_-, Q_+\} = -  \frac{\cal Z}{\sqrt 2} \,,
 }
which, up to the normalization of $Q_\pm$ and of the central charge ${\cal Z}$, is the standard ${\cal N} = (1, 1)$ supersymmetry algebra (for a review, see \cite{Shifman:2007ce}).  The same algebra, but without the central term, was obtained in the light-cone approach in \cite{Boorstein:1993nd} directly from the supercharges, without working with the corresponding supercurrents.
We should note, however, that since the supersymmetric Adjoint QCD$_2$ model (\ref{STotal}) has unique vacuum, there is no possibility of having domain walls; therefore, the central term vanishes.\footnote{In Section~\ref{MASSLESS}, we will generalize the model by adding massless fermionic fields, so that multiple degenerate vacua become possible. In these cases, there can be domain walls interpolating between these vacua, but the central term still appears to vanish.  In the cases that we studied, ${\cal Z}$ vanishes because the degenerate vacua are related by the action of a non-invertible symmetry whose generators commute with $Z$.  Thus, $\langle Z \rangle$ will take the same value in all such vacua, and consequently ${\cal Z}$ will vanish.\label{VanishingFootnote}}

\subsection{Generalization to other gauge groups}

The construction in the previous four subsections can be generalized by replacing $\SU(N)$ with an arbitrary simply-connected gauge group $G$, with only few other very minor modifications.  First, the trace, which for $\SU(N)$ was defined to be in the fundamental representation, should be replaced with the trace in the adjoint representation, rescaled by an appropriate factor:
 \es{trReplacement}{
  \tr \to \frac{1}{2 h^\vee} \tr_\text{adj} \,,
 }
where $h^\vee$ is the dual Coxeter number, and the trace in the adjoint representation is normalized such that $\tr_\text{adj} (T^a T^b) = h^\vee \delta^{ab}$.  Second, every instance of $N$ should be replaced with the dual Coxeter number
 \es{Nreplacement}{
  N \to h^\vee \,.
 }
Thus, the theory with arbitrary $G$ is ${\cal N} = (1, 1)$ supersymmetric with supercurrent
 \es{LCsupercurrentsGen}{
  j_{--} &= \frac{1}{6} \psi_-^a J_-^a \,,  \qquad j_{+-} = \frac{\alpha}{2} \psi_+^a F^a \,, \\
  j_{++} &= \frac{1}{6} \psi_+^a J_+^a  \,, \qquad  j_{-+} =\frac{\alpha}{2}  \psi_-^a F^a  \,,
 }
provided that
 \es{normalizationsGen}{
  m =  \sqrt{\frac{g^2 h^\vee}{2\pi}} \,,
    \qquad 
    \alpha = \sqrt{\frac{h^\vee} {4 g^2 \pi}} \,.
 }

\section{Adding massless fermions}
\label{MASSLESS}

 \subsection{Setup}

Let us see how this construction of the previous section generalizes to the $\SU(N)$ gauge theory containing one massive adjoint fermion and additional massless fermions $q_\alpha$ in some (not necessarily irreducible) representation ${\cal R}$, with $\alpha$ being an index for the states of ${\cal R}$.  The combined action is  
 \es{SFermions}{
 S = \int d^2x\, \left[ \tr\left(-\frac{1}{2 g^2}F_{\mu\nu}F^{\mu\nu} + i
\overline{\psi}\slashed{D}\psi - m  \overline{\psi}\psi\right) +  i \overline{q}_\alpha\slashed{D}q_\alpha \right]  \, ,
 }
where $D_\mu q_{\alpha} = \partial_\mu q_{\alpha} - i A_\mu^a t_{\alpha \beta}^a q_{\beta}$, with $t^a_{\alpha\beta}$ being the $\SU(N)$ generators in representation ${\cal R}$.\footnote{If ${\cal R}$ is real, we should include an extra factor of $1/2$ in the last term in \eqref{SFermions}, as well as in the definitions of $\widetilde{J}_\mu$ in \eqref{ModCurrents}, and in the corresponding terms in the definition of the stress tensor in \eqref{GotT}.}  We again assume $m \geq 0$ without loss of generality, since one can flip the sign of $m$ by sending $\psi \to \gamma^5 \psi$.  

In \eqref{SFermions}, the gauge field couples to the total current
 \es{Jtot}{
  J^\text{tot}_\mu = J_\mu + \widetilde{J}_\mu \,,
 } 
where $J_\mu = \overline{\psi} \gamma_\mu \psi$ as before, and $\widetilde{J}_\mu^a = -\overline{q}_\alpha \gamma_\mu t^a_{\alpha\beta} q_\beta$.  Upon defining $q_{\alpha} = 2^{-1/4}\begin{pmatrix} q_{\alpha -} \\ q_{\alpha +}\end{pmatrix}$, we can write the gauge currents as:
 \es{ModCurrents}{
  J_-^\text{tot} &= J_- + \widetilde{J}_- \,, \qquad J_-^a =  \frac{i}{2} f^{abc} \psi_-^b \psi_-^c \,, \qquad  
    \widetilde{J}_-^{a}  = - q_{\alpha -}^\dagger  t_{\alpha \beta }^a q_{\beta -}  \,, \\
 J_+^\text{tot} &= J_+ + \widetilde{J}_+ \,, \qquad J_+^a =  \frac{i}{2} f^{abc} \psi_+^b \psi_+^c \,, \qquad  
    \widetilde{J}_+^{a}  = - q_{\alpha +}^\dagger  t_{\alpha \beta }^a q_{\beta +}   \,.
 }
The gauge currents should be interpreted as being appropriately regularized by point splitting in a gauge covariant way.  For instance,
 \es{JPointSplitting}{
  \widetilde{J}_\mu^a(x) = - \lim_{\epsilon \to 0}  \overline{q}_\alpha(x) \gamma_\mu t^a_{\alpha\beta} U_{\beta \gamma}(x, x+ \epsilon) q_\gamma(x + \epsilon) \,,
 }
where $U_{\beta \gamma}(x_1, x_2)$ is the parallel transport operator \eqref{UDef} evaluated in representation ${\cal R}$.

The equations of motion following from action \eqref{SFermions} are
 \es{eomsGen}{
    J_-^\text{tot} &= \frac{1}{g^2}D_- F \,, \qquad  \ \, J_+^\text{tot} = -\frac{1}{g^2} D_+ F \,, \\
     D_-\psi_+ &= \frac{m}{\sqrt{2}} \psi_- \,, \qquad   D_+\psi_- = -\frac{m}{\sqrt{2}} \psi_+ \,, \\
    D_- q_{\alpha +} &= 0 \,, \qquad \qquad    D_+ q_{\alpha -} = 0 \,.
 }
The currents $\widetilde{J}_\mu$ form an $\SU(N)_k$ current algebra
 \es{CurrentOPE}{
  \widetilde{J}^a_-(x) \widetilde{J}^b_-(0) &= - \frac{k \delta^{ab}}{8 \pi^2 (x^-)^2} - \frac{f^{abc}}{2 \pi x^-} \widetilde{J}_-^c + \cdots \,, \\
  \widetilde{J}^a_+(x) \widetilde{J}^b_+(0) &= - \frac{k \delta^{ab}}{8 \pi^2 (x^+)^2} - \frac{f^{abc}}{2 \pi x^+} \widetilde{J}_+^c + \cdots \,,
 }
where the level $k$ of the algebra is 
 \es{Gotk}{
  k =  \frac{C_2({\cal R})\, \text{dim}\, {\cal R}}{N^2 - 1}  \,,
 } 
where $C_2({\cal R})$ is the quadratic Casimir, and $\text{dim}\, {\cal R}$ equals the (real) dimension of ${\cal R}$ if ${\cal R}$ is real and twice the complex dimension of ${\cal R}$ if ${\cal R}$ is complex.   For example, if ${\cal R}$ is $N_f$ copies of the fundamental representation, then $k = N_f$;  if ${\cal R}$ is the adjoint representation, then $k = N$.  One can show that if the currents $\widetilde{J}_\mu^a$ are regularized in a gauge-invariant way, as in \eqref{JPointSplitting}, the space-time derivatives acquire terms proportional to the gauge field strength $F = F_{+-}$, just as in \eqref{DJp}--\eqref{DJm} with $m \to 0$ and $N \to k$
\es{DtildeJ}{
  D_+ \widetilde{J}_- =  - \frac{k}{4 \pi} F \,,  \qquad D_- \widetilde{J}_+ =   \frac{k}{4 \pi} F \,,
 }
ensuring that $D^\mu \widetilde{J}_\mu = D_+ \widetilde{J}_- + D_- \widetilde{J}_+ = 0$.  The derivation of \eqref{DtildeJ} is similar to that in Appendix~\ref{EQUALTIME} for the case of adjoint fermions.

This theory has the canonical stress tensor (by ``canonical'' we mean the stress tensor used to couple this theory to gravity) 
 \es{GotT}{
   \begin{aligned} T_{--} &=  i \tr \left(\psi_- D_- \psi_- \right) + i q_{\alpha -}^\dagger D_- q_{\alpha -}  \,, \\
  T_{+-} & =  \frac{1}{g^2} \tr F^2 + \frac{i m}{\sqrt{2}} \tr (\psi_+ \psi_-) = T_{-+}  \,, \\
  T_{++} &= i \tr \left(\psi_+ D_- \psi_+ \right) + i q_{\alpha +}^\dagger D_+ q_{\alpha +} \,.
  \end{aligned}
 }
As we will show, the theory defined this way is (partially) supersymmetric at a certain value of $m$ that was originally determined in \cite{Popov:2022vud}:
\es{SUSYMassModified}{
  m = \sqrt{\frac{g^2 (N+k)}{2\pi}} \,.
 }

\subsection{Supercurrent}

To find the supercurrent, we start with an ansatz where two of the components pick up contributions from $\widetilde{J}_\mu$ while the other two have the same form as in \eqref{LCsupercurrents}
 \es{ModifiedSupercurrents}{
   j_{--} &=  \frac{1}{3} \tr \psi_-^3 + \tr (\psi_- \widetilde{J}_- ) \,,
       \qquad j_{+-} = \alpha \tr \left( \psi_+ F \right) \,, \\
   j_{++} &= \frac{1}{3} \tr \psi_+^3 + \tr (\psi_+ \widetilde{J}_+) \,, 
       \qquad  j_{-+} = \alpha \tr \left( \psi_- F \right) \,.
 }
We will find that the supercurrent is conserved for mass (\ref{SUSYMassModified}) and
 \es{AGen}{
  \alpha =  \sqrt{\frac{N+k} {4 g^2 \pi}} \,.
 }

We can check the conservation equation $\partial_- j_{+-} + \partial_+ j_{--} = 0$ as follows.  First, the equations of motion \eqref{eomsGen} imply
 \es{dmjpGen}{
   \partial_- j_{+-} = \alpha \frac{m}{\sqrt{2}}\tr ( \psi_-   F) + \alpha g^2 \tr \left( \psi_+ J^\text{tot}_-  \right) \,.
 }
For computing $\partial_+ j_{--}$, we already have from \eqref{djQuantum} that 
 \es{dpjmGen1}{
  \partial_+ \frac 13 \tr \psi_-^3 = - \frac{m}{\sqrt{2}} \tr \left( \psi_+ J_- \right)  - \frac{N}{4 \pi} \tr \left( \psi_- F \right) \,.
 }
But now $j_{--}$ has an extra term $\tr (\psi_- \widetilde{J}_- )$.  Its derivative can be computed using the equation of motion for $\psi_-$ in \eqref{eomsGen} and the anomaly equation \eqref{DtildeJ}:
 \es{dpjmGen2}{
  \partial_+ \tr (\psi_- \widetilde{J}_- ) = - \frac{m}{\sqrt{2}} \tr (\psi_+ \widetilde{J}_-) - \frac{k}{4 \pi} \tr (\psi_- F) \,.
 }
Adding together \eqref{dpjmGen1} and \eqref{dpjmGen2}, we find
 \es{dpjmGen}{
  \partial_+ j_{--}  =  - \frac{m}{\sqrt{2}} \tr \left(\psi_+ J_-^\text{tot} \right) - \frac{N+k}{4\pi} \tr (\psi_- F) \,.
 }
The current conservation $\partial_- j_{+-} + \partial_+ j_{--} = 0$ is again satisfied provided that $\alpha$ and $m$ are given in \eqref{AGen} and \eqref{SUSYMassModified}, respectively.   The conservation of the right-chiral component of the supercurrent can be checked analogously.

The conserved supercharges are modified to
\es{GotQEqualTimeGen}{
  Q_- &= \int dx\, \frac{1}{\sqrt{2}} \left[ \frac{1}3 \tr \left( \psi_-^3  \right)  + \tr (\psi_- \widetilde{J}_-) +  \alpha \tr (F \psi_+)  \right] \,, \\
  Q_+ &= \int dx\, \frac{1}{\sqrt{2}}  \left[   \frac {1}3 \tr \left(\psi_+^3 \right)  + \tr (\psi_+ \widetilde{J}_+) + \alpha \tr (F \psi_- ) \right]  \,.
 }
In light-cone quantization, the supercharges take the simpler forms $Q_- =  \int dx^-\, j_{--} = \int dx^- \, \left[ \frac 13 \tr \psi_-^3 + \tr (\psi_- \widetilde{J}_-) \right]$ and $Q_+ = \int dx^- \, j_{-+} = \alpha \int dx^- \, \tr (\psi_- F)$.  These expressions agree with those found in \cite{Popov:2022vud} in the gauge $A_-= 0$ after eliminating $A_+$ and $\psi_+$ using their equations of motion.

\subsection{The supercurrent multiplet}

The supercurrent multiplet has the same structure as discussed in Section~\ref{MULTIPLET} in the adjoint QCD$_2$ theory.  It contains a scalar operator $Z$, the supercurrent, and a conserved stress tensor that, as it will turn out, differs from the canonical stress tensor.  To compute the required commutation relations we again use the anti-commutators and commutators in \eqref{CanRels} and \eqref{FAdditional}, as well as
 \es{FAdditional2}{
  \left[F^a(t, x), \tr (\psi_- \widetilde{J}_-)(t, y)\right] &=   \frac{ig^2 k}{4\sqrt{2}\pi}   \delta(x-y)  \psi_-^a(t, y) \,, \\
  \left[F^a(t, x), \tr (\psi_+ \widetilde{J}_+)(t, y)\right] &=   -\frac{ig^2 k}{4\sqrt{2}\pi}   \delta(x-y)  \psi_+^a(t, y) \,,
 }
which arise as a consequence of the gauge-covariant point-splitting regularization of the currents $\widetilde{J}_\mu^a$ in a way similar to that described in Appendix~\ref{EQUALTIME}.  Many of the (anti-)commutation relations we need, given explicitly in \eqref{RelevantComm} and \eqref{RelevantAnticomm}, were already used before.  The only new ones are those in \eqref{RelevantCommGen} and \eqref{RelevantAnticommGen}.

Let us uncover the structure of the multiplet.  We start with the scalar operator $Z$ that takes the same form as \eqref{ZDef}, 
 \es{ZDefAgain}{
  Z = \frac{\alpha}{g^2} \tr F^2 \,,
 }
but now with the modified value of $\alpha$ in \eqref{AGen}.  We find that its commutators with the supercharges \eqref{GotQEqualTimeGen} give two of the components of the supercurrent:
 \es{QOGen}{
  [Q_-, Z] = - \frac{i (N+k)}{8 \pi} j_{-+} \,, \qquad 
   [Q_+, Z] = - \frac{i (N+k)}{8 \pi} j_{+-} \,,
 }
as expected.  The derivatives of $Z$ can also be obtained from anti-commutators of the mixed chirality components of $Q$ and $j$:
 \es{AntiComm2}{
  \{Q_-, j_{-+}\} = \frac 12 \partial_- Z  \,, \qquad \{Q_-, j_{++}\} = -\frac 12 \partial_+ Z \,, \\
  \{Q_+, j_{--}\} = \frac 12 \partial_- Z  \,, \qquad \{Q_+, j_{+-}\} = -\frac 12 \partial_+ Z \,.
 }
The supersymmetric stress tensor can be computed via the anti-commutators of the same chirality components of $Q$ and $j$:
 \es{AntiComm3}{
  \{ Q_-, j_{--} \} &= \frac{N+k}{8 \pi} T^\text{SUSY}_{--} \,, \qquad
   \{ Q_-, j_{+-}\} =  \frac{N+k}{8 \pi} T^\text{SUSY}_{+-} \,, \\
  \{ Q_+, j_{++} \} &=  \frac{N+k}{8 \pi} T^\text{SUSY}_{++} \,, \qquad
   \{ Q_+, j_{-+}\} =  \frac{N+k}{8 \pi} T^\text{SUSY}_{-+} \,,
 }
where the overall factor of $\frac{N+k}{8 \pi}$ was chosen such that $T^\text{SUSY}_{\mu\nu}$ ends up being canonically normalized in gapped theories.  Using \eqref{RelevantComm}--\eqref{RelevantAnticomm} and \eqref{RelevantCommGen}--\eqref{RelevantAnticommGen}, we find
 \es{GotTildeT}{
   T^\text{SUSY}_{--} &=  i \tr \left(\psi_- D_- \psi_- \right) + \frac{4 \pi}{N+k} \tr \widetilde{J}_-^2 \,, \\
  T^\text{SUSY}_{+-} & =  \frac{1}{g^2} \tr F^2 + \frac{i m}{\sqrt{2}} \tr (\psi_+ \psi_-) = T^\text{SUSY}_{-+}  \,, \\
  T^\text{SUSY}_{++} &= i \tr \left(\psi_+ D_- \psi_+ \right) + \frac{4 \pi}{N+k} \tr \widetilde{J}_+^2 \,.
 }
This differs from the canonical stress tensor in that the part coming from the current algebra degrees of freedom are replaced by an appropriately regularized (by point splitting, with a small Wilson line inserted between the two points) gauge-invariant version of the Sugawara stress tensor $T^\text{Sugawara}_{\mu\nu}$ for the $\SU(N)_k$ currents $\widetilde{J}_\mu$.   Here,
 \es{TSugawara}{
  T^\text{Sugawara}_{--} =  \frac{4 \pi}{N+k} \tr \widetilde{J}_-^2 \,, \qquad T^\text{Sugawara}_{++} =  \frac{4 \pi}{N+k} \tr \widetilde{J}_+^2 \,, \qquad T^\text{Sugawara}_{\pm \mp}= 0 \,.
 }
The Sugawara stress tensor generally contains four-fermion terms that would not be present in the canonical stress tensor.  We discuss this in more detail below.

The SUSY algebra reads
 \es{SUSYAlgGen}{
  \{ Q_-, Q_-\} = \frac{N+k}{8 \pi}  P^\text{SUSY}_- \,, \qquad \{Q_+, Q_+\} = \frac{N+ k}{8 \pi}  P^\text{SUSY}_+ \,, \qquad \{Q_-, Q_+\} = - \frac{\cal Z}{\sqrt 2} \,,
 }
where $P^\text{SUSY}_\mu = \int dx \, (T^\text{SUSY})^0{}_\mu$ and the central charge ${\cal Z}$ has the same expression as in \eqref{calZDef}. As noted earlier (see Footnote~\ref{VanishingFootnote}), there appear to be no domain wall configurations for which the central charge ${\cal Z}$ is non-vanishing.

\subsection{Conservation of the supersymmetric stress tensor}

The supersymmetric stress tensor is conserved just by virtue of being in the same multiplet as the conserved supercurrent.  Indeed, in a more covariant notation, we can write the commutators and anti-commutators with $Q_\pm$ in \eqref{QOGen}, \eqref{AntiComm2}, and \eqref{AntiComm3} as
 \es{SUSYTransf2Gen}{
  [Q, Z ] = \frac{N+k}{8 \sqrt{2} \pi} \gamma^\nu j_\nu \,, \qquad
   \{ j_\mu, \overline{Q} \} = \frac{N+k}{16 \pi} T_{\mu\nu}^\text{SUSY} \gamma^\nu - \frac{i}{2 \sqrt{2}} \gamma^5 \epsilon_{\mu\rho} \partial^\rho Z \,.
 } 
It is easy to see that $\partial^\mu j_\mu = 0$ implies the conservation condition $\partial^\mu T_{\mu\nu}^\text{SUSY} = 0$.

It is insightful, however, to also check the conservation explicitly from \eqref{GotTildeT}.  Let us check just the minus component of the conservation equation,
 \es{ConsTSUSY}{
  \partial_+ T_{--}^\text{SUSY} + \partial_- T_{+-}^\text{SUSY} = 0 \,,
 }
the check of the other component being performed analogously.  For $\partial_- T_{+-}^\text{SUSY}$, we find
 \es{dmTpm}{
  \partial_- T^\text{SUSY}_{+-} = 2 \tr (F J^\text{tot}_- ) + \frac{i m}{\sqrt{2}} \tr (\psi_+ D_- \psi_-)  \,,
 }
where the first term was obtained using the Maxwell equation $\partial_- F = g^2 J_-^\text{tot}$.  The computation of $ \partial_+ T_{--}^\text{SUSY}$ is more subtle.    The derivative of the first term in $T_{--}^\text{SUSY}$ in \eqref{GotTildeT} is again straightforward
 \es{dpTmm1}{
  \partial_+ \left(  i \tr \left(\psi_- D_- \psi_- \right) \right) =  - \frac{i m}{\sqrt{2}} \tr (\psi_+ D_- \psi_-) - 2 \tr (F \psi_- \psi_- ) \,, 
 }
where the first term is obtained by taking the derivative of $\psi_-$ in $\tr \left(\psi_- D_- \psi_- \right)$ using $D_+ \psi_- = - \frac{m}{\sqrt{2}} \psi_+$, and the second term was obtained from $D_+ D_- \psi_- = [D_+, D_-] \psi_- + D_- D_+ \psi_- = - i [F, \psi_-] - \frac{m}{\sqrt{2}} D_- \psi_+$ and the equation of motion $D_- \psi_+ = \frac{m}{\sqrt{2}} \psi_-$. 
For the derivative of the Sugawara term $\frac{4 \pi}{N+k} \tr \widetilde{J}_-^2$ we can use the relation \eqref{DerAB2} for derivatives of composite operators.  If we take ${\cal A} = {\cal B} = \widetilde{J}$, the current algebra \eqref{CurrentOPE} gives\footnote{The leading term in the OPE \eqref{CurrentOPE} would give a contribution to $\partial_+ \tr  \widetilde{J}_-^2$ proportional to $k f^{aba} F^b$, which trivially vanishes due to the anti-symmetry of the structure constants.} ${\cal C} = - \frac{1}{2 \pi} \widetilde{J}_-$, and \eqref{DerAB2} becomes
 \es{DerSug}{
  \partial_+ \tr  \widetilde{J}_-^2 = \tr \left( (D_+ \widetilde{J}_-) \widetilde{J}_- \right) 
   + \tr \left( \widetilde{J}_- (D_+ \widetilde{J}_-) \right)  - \frac{N}{2 \pi} \tr ( F \widetilde{J}_- )
    =  - \frac{N + k}{2 \pi} \tr ( F \widetilde{J}_-  )  \,,
 }
where in the last equality we also used \eqref{DtildeJ}.  Adding \eqref{dpTmm1} to \eqref{DerSug} multiplied by 
$\frac{4 \pi}{N+k}$ gives
 \es{dpTmm}{
  \partial_+ T^\text{SUSY}_{--} =  - \frac{i m}{\sqrt{2}} \tr (\psi_+ D_- \psi_-)  - 2 \tr (F J^\text{tot}_- )\,.
 }
 The conservation equation \eqref{ConsTSUSY} follows.
 
 The lesson that can be learned from this calculation is that for $T_{\mu\nu}^\text{SUSY}$ to be conserved, we must use both a gauge-covariant definition of $\widetilde{J}_\mu$ (such that the anomaly equations \eqref{DtildeJ} hold) as well as a gauge-invariant definition of the products of two currents $\tr (\widetilde{J}_-^2)$ and $\tr (\widetilde{J}_+^2)$.

\subsection{Partial supersymmetry}
\label{PARTIAL}

The canonical stress tensor \eqref{GotT} can be written as
 \es{DeltaTDef}{
 T_{\mu\nu}  = T^\text{SUSY}_{\mu\nu} +  \Delta T_{\mu\nu} \,,
 }
with
 \es{GotDeltaT}{
  \Delta T_{--} &=  i q_{\alpha -}^\dagger D_- q_{\alpha -} - \frac{4 \pi}{N+ k} \tr \widetilde{J}_-^2\,, \\
  \Delta T_{+-} &= \Delta T_{-+} = 0 \,, \\
  \Delta T_{++} &=   i q_{\alpha +}^\dagger D_+ q_{\alpha +} - \frac{4 \pi}{N+ k} \tr \widetilde{J}_+^2 \,.
 } 
Since $\Delta T_{\mu\nu}$ is the difference of two conserved stress tensors, it is also a conserved stress tensor.  Moreover, it is straightforward to check that the components of $\Delta T_{\mu\nu}$ commute with the components of $T^\text{SUSY}_{\mu\nu}$, namely $ [\Delta T_{\mu\nu}, T^\text{SUSY}_{\rho \sigma} ] = 0 $ for any $\mu$, $\nu$, $\rho$, $\sigma$.  Thus, $T^\text{SUSY}_{\mu\nu}$ and $\Delta T_{\mu\nu}$ are the stress tensors of two decoupled sectors of the theory, only the sector corresponding to $T^\text{SUSY}_{\mu\nu}$ being supersymmetric.

There is one subtlety in checking that $\Delta T_{\mu\nu}$ commutes with $T^\text{SUSY}_{\mu\nu}$.  Naively, one may think it does not commute with the $\tr F^2$ term in $T^\text{SUSY}_{+-}$ since the gauge field appears explicitly in the covariant derivatives in \eqref{GotDeltaT} and through the regularization of the operator $\tr \widetilde{J}_\pm^2$.  However, these two effects precisely cancel, and one can show that
 \es{FCommDeltaT}{
  [ F^a(t, x), \Delta T_{\mu\nu}(t, y) ] = 0 \,. 
 }
This equation implies that the formulas in \eqref{GotDeltaT}, with $\widetilde{J}_\pm^2$ defined by gauge-invariant point-splitting, are equivalent to what one would get if one replaces the covariant derivatives by ordinary ones in \eqref{GotDeltaT} and also interprets $\tr \widetilde{J}_\pm^2$ as a normal-ordered product of the two currents.  Thus, we can also write $\Delta T_{\mu\nu}$ as
  \es{GotDeltaTAgain}{
  \Delta T_{--} &=  i q_{\alpha -}^\dagger \partial_- q_{\alpha -} - \frac{4 \pi}{N+ k} \tr \widetilde{J}_-^2\,, \\
  \Delta T_{+-} &= \Delta T_{-+} = 0 \,, \\
  \Delta T_{++} &=   i q_{\alpha +}^\dagger \partial_+ q_{\alpha +} - \frac{4 \pi}{N+ k} \tr \widetilde{J}_+^2  \,,
 } 
where $ \tr \widetilde{J}_\pm^2$ are now regularized in the ungauged theory of the $q_\alpha$ fermions.

The stress tensor $\Delta T_{\mu\nu}$ in \eqref{GotDeltaTAgain} is precisely the stress tensor of the coset CFT
 \es{CosetCFT}{
  \frac{\SO(\dim {\cal R})_1}{ \SU(N)_k} \,,
 }
where $\dim {\cal R}$ represents the number of Majorana fermions.  In general, this coset CFT is non-supersymmetric and it has Virasoro central charge $c_\text{IR} = c_{\SO(\dim {\cal R})_1} - c_{\SU(N)_k}$.    It was shown in \cite{Goddard:1985jp} that $\Delta T_{\mu\nu} = 0$ if and only if $c_\text{IR} = 0$ (see also \cite{Delmastro:2021otj}), so in such a case $T_{\mu\nu} = T_{\mu\nu}^\text{SUSY}$ and the whole theory is supersymmetric.

While in general $\Delta T_{\mu\nu}$ describes a gapless sector, the supersymmetric stress tensor $T^\text{SUSY}_{\mu\nu}$ describes a gapped sector in the trivial flux tube sector.  Indeed, in the deep infrared the gauged current is $J^\text{tot}_\mu = 0$, and the massive fermion can be integrated out.  See also the discussion in Section 4.2 of \cite{Delmastro:2021otj}.

\subsection{A fully supersymmetric model with two adjoints}
\label{TwoAdjoint}

As a particular case, let us consider the $\SU(N)$ gauge theory coupled to a massive adjoint $\psi$ and a massless adjoint $\widetilde{\psi}$,
\es{Stwoadj}{
    S  = \int d^2x \tr\left(-\frac{1}{2 g^2}F_{\mu\nu}F^{\mu\nu} + i
\overline{\psi}\slashed{D}\psi +i \overline{\widetilde\psi}\slashed{D}\widetilde\psi    - m  \overline{\psi}\psi\right) \, .
} 
This model has $\SU(N)$ current algebra of level
 $k=N$, so it becomes fully supersymmetric for $m= \sqrt\frac{g^2 N}{\pi}$. The explicit expression for the supercurrent is
\es{TwoadjSupercurrents}{
   j_{--} &=  \frac{1}{3} \tr \psi_-^3 + \tr ( \psi_- \widetilde{\psi}_- \widetilde{\psi}_- )  \,,
       \qquad j_{+-} = \sqrt{\frac{N} {2 g^2 \pi}} \tr \left( \psi_+ F \right) \,, \\
   j_{++} &= \frac{1}{3} \tr \psi_+^3 + \tr ( \psi_+ \widetilde{\psi}_+ \widetilde{\psi}_+ ) \,, 
       \qquad  j_{-+} = \sqrt{\frac{N} {2 g^2 \pi}} \tr \left( \psi_- F \right) \,.
 }
This model has $c_{\rm IR}=0$ so that the supersymmetry should be exact. Indeed, it is not hard to check that the extraneous four-fermion terms in $P_\pm^{\text{SUSY}}$ cancel. Some of the cancellations are obvious, since $ \tr \psi_-^4=\tr \widetilde \psi_-^4 $ vanish trivially.

Unlike the supersymmetric model with a single massive adjoint, the supersymmetric model with two adjoints has a vanishing string tension. Therefore, there is no spontaneous supersymmetry breaking, and all $2^{N-1}$ degenerate vacua are gapped.  These vacua are described by the topological coset model $\frac{\SO(N^2-1)_1}{\SU(N)_{N}}$, just like for the non-supersymmetric model with a single massless adjoint. 
Numerical Discretized Light-Cone Quantization studies of the supersymmetric model with two adjoints were started in \cite{AlexMcDonaldJP}, and it would be interesting to study this model further.

\subsection{Generalization to other gauge groups}
\label{GENERALGAUGE}

As in the basic adjoint QCD${}_2$ model, the construction in the previous six subsections can be generalized by replacing $\SU(N)$ with an arbitrary simply-connected gauge group $G$, with the same modifications as in \eqref{trReplacement} and \eqref{Nreplacement}.   
The current algebra for the massless fermions is as in \eqref{CurrentOPE}, with the level $k$ being
 \es{GotkGen}{
  k =  \frac{C_2({\cal R})\, \text{dim}\, {\cal R}}{\text{dim}\, G}  \,,
 } 
where $C_2({\cal R})$ is the quadratic Casimir, $\text{dim}\, {\cal R}$ is the real dimension of ${\cal R}$ and $\text{dim}\, G$ is the dimension of $G$.  
The supercurrent 
 \es{LCsupercurrentsGen2}{
  j_{--} &= \frac{1}{6} \psi_-^a J_-^a  + \frac 12 \psi_-^a \widetilde{J}_-^a \,,  \qquad j_{+-} = \frac{\alpha}{2} \psi_+^a F^a \,, \\
  j_{++} &= \frac{1}{6} \psi_+^a J_+^a + \frac 12 \psi_+^a \widetilde{J}_+^a \,, \qquad  j_{-+} =\frac{\alpha}{2}  \psi_-^a F^a 
 }
is conserved provided that
 \es{normalizationsGen2}{
  m =  \sqrt{\frac{g^2 (h^\vee+k)}{2\pi}} \,,
    \qquad 
    \alpha =  \sqrt{\frac{h^\vee+k} {4 g^2 \pi}} \,.
 }

\subsection{A $\grU(1)$ example}
\label{Abelian}

The generalization to arbitrary $G$ of the previous subsection also works when $G = \grU(1)$.  Indeed, a simple manifestly supersymmetric $\grU(1)$ gauge theory was discussed in \cite{Popov:2022vud} as a warm-up example. It is the one-flavor massless Schwinger model \cite{Schwinger:1962tp} with an added free neutral Majorana fermion $\psi$ of mass 
$m=\frac{g}{\sqrt \pi}$.   Denoting the massless Dirac fermion of the Schwinger model by $q$, the action takes the form\footnote{In the $\grU(1)$ case we have a single Lie algebra generator, so we omit the Lie algebra index.}
 \es{SSchwinger}{
   S = \int d^2x\, \left[ -\frac{1}{4 g^2}F_{\mu\nu}F^{\mu\nu} +  i \overline{q} (\slashed{\partial} - i \slashed{A}) q 
     +  \frac i2 \overline{\psi} \slashed{\partial}\psi - \frac{g}{2 \sqrt{\pi}}  \overline{\psi}\psi \right]  \,.
 }
  In this case $J_\mu = 0$ since a Majorana fermion is neutral under the $\grU(1)$ gauge group, and $\widetilde J_\mu = - \overline{q} \gamma_\mu q$.  In the normalization convention we have been using, the level of the current algebra \eqref{GotkGen} is $k=2$ and $h^\vee = 0$.

Since the spectrum of massless Schwinger model consists of a free real scalar field, which also has mass $\frac{g}{\sqrt \pi}$, this model has manifest $(1,1)$ supersymmetry. While this model may seem trivial, it is instructive to write down its supercurrent using the fermionic fields. We will find that its conservation is a consequence of the famous chiral anomaly equation
\cite{Johnson:1963vz,Peskin:1995ev}
 in the Schwinger model:
 \es{ChiralAnomSchwinger}{
  \partial^\mu \widetilde{J}_\mu{}^5 = -\frac{1}{\pi} F \,,
 }
where $\widetilde{J}_\mu{}^5 = - \overline{q} \gamma_\mu \gamma^5 q$ is the axial current.

The expression for the supercurrent takes the form\footnote{When writing \eqref{Schwingersupercurrents} and \eqref{Schwingersupercurrents2} below, we rescaled \eqref{ModifiedSupercurrents} by a factor of $2$ for convenience.}
 \es{Schwingersupercurrents}{
  j_{--} &= - \psi_- q_-^\dagger q_-  = \psi_- \widetilde{J}_- \,,  \qquad j_{+-} = \frac{1}{g \sqrt{2\pi}}  \psi_+ F  \,, \\
  j_{++} &= - \psi_+ q_+^\dagger q_+   = \psi_+ \widetilde{J}_+ \,, \qquad  j_{-+} = \frac{1}{g \sqrt{2\pi}}  \psi_- F  \,.
 }
The conservation of $\widetilde{J}_\mu$ and the Schwinger anomaly equation imply
  \es{DerJU1}{
\partial_- \widetilde{J}_+ = \partial_+ \widetilde{J}_- = -\frac{1}{2\pi} F \,,
 }
as in \eqref{DtildeJ} with $k=2$.   Using \eqref{DerJU1}, we readily find that the supercurrent is conserved.

This construction can be generalized to $\grU(1)$ gauge theory with $N_f>1$ massless charged Dirac fermions $q_\alpha$, $\alpha=1, \ldots, N_f$, and a massive neutral Majorana fermion of mass $\frac{g\sqrt{N_f}}{\sqrt \pi}$:
 \es{SSchwinger2}{
   S = \int d^2x\, \left[ -\frac{1}{4 g^2}F_{\mu\nu}F^{\mu\nu} +  \sum_{\alpha = 1}^{N_f} i \overline{q}_\alpha (\slashed{\partial} - i \slashed{A}) q_\alpha 
     +  \frac i2 \overline{\psi} \slashed{\partial}\psi - \frac{g \sqrt{N_f} }{2 \sqrt{\pi}}  \overline{\psi}\psi \right]  \,.
 }
Then the conserved supercurrent is
\es{Schwingersupercurrents2}{
  j_{--} &= - \psi_- \sum_{\alpha=1}^{N_f} q_{\alpha-}^{\dagger} q_{\alpha-}  = \psi_- \widetilde{J}_- \,,  \qquad j_{+-} = \frac{1}{g \sqrt {2 \pi N_f}}  \psi_+ F  \,, \\
  j_{++} &= - \psi_+ \sum_{\alpha=1}^{N_f} q_{\alpha+}^{\dagger} q_{\alpha+}     = \psi_+ \widetilde{J}_+ \,, \qquad  j_{-+} = \frac{1}{g \sqrt {2 \pi N_f}} \psi_- F  \,.
 }
This supercurrent describes the free supersymmetric theory consisting of a real scalar and a Majorana fermion of the same mass $\frac{g\sqrt{N_f}}{\sqrt \pi}$.
The model also contains a decoupled non-supersymmetric CFT sector described by the $\grSU(N_f)_1$ WZW model \cite{Gepner:1984au,Affleck:1985wa}. 

\subsection{Fully supersymmetric gapless theories}
\label{ThreeAdjoint}

If the coset CFT sector has $c_{\rm IR}>0$, then it is typically not supersymmetric. In such cases, the massive sector is still supersymmetric so that the theory possesses partial supersymmetry. However, for the special choices of massless fermionic matter where the CFT sector is supersymmetric, the theory may be fully supersymmetric. 

In this section we present an example of a family of such theories.  We start with the $\SU(N)$ gauge theory coupled to one adjoint $\psi$ of mass $m= \sqrt\frac{3 g^2 N}{2\pi}$ and {\it two} massless adjoints $\widetilde{\psi}_i$, where $i=1,2$.  The low-energy limit of this theory is described by the coset CFT $\frac{\SO(2N^2-2)_1}{\SU(N)_{2N}}$ of Virasoro central charge $\frac{N^2-1}{3}$, which has ${\cal N}=(2, 2)$ supersymmetry \cite{Gopakumar:2012gd,Isachenkov:2014zua,Damia:2024kyt}. Let us show that the full gauge theory has
${\cal N}=(1,1)$ supersymmetry. 

For the massive sector, we have our usual supercurrent \eqref{ModifiedSupercurrents} coupling the massless and massive fields,
\es{MassiveSupercurrents}{
   j_{--} &=  \frac{1}{3} \tr \psi_-^3 + \tr ( \psi_- \widetilde{\psi}_{i-} \widetilde{\psi}_{i-} )  \,,
       \qquad j_{+-} =  \alpha \tr \left( \psi_+ F \right) \,, \\
   j_{++} &= \frac{1}{3} \tr \psi_+^3 + \tr ( \psi_+ \widetilde{\psi}_{i+} \widetilde{\psi}_{i+} ) \,, 
       \qquad  j_{-+} = \alpha \tr \left( \psi_- F \right) \,.
 }
It is conserved provided that $\alpha= \sqrt{\frac{3 N} {4 g^2 \pi}}$, since in this case the level of the current algebra for the massless fermions is $k = 2N$. The anti-commutators of the supercurrent with the supercharges produces the stress energy tensor of the massive sector as in \eqref{AntiComm3}, with coefficient $3N / (8 \pi)$.

For the coset CFT, ${\cal N} = (2, 2)$ supersymmetry implies that there exist two linearly-independent supercurrents $j_{\mu A i}^\text{CFT}$, with $i = 1, 2$, and two corresponding supercharges
 \es{QDef}{
  Q_{Ai}^\text{CFT} = \int dx\, (j^\text{CFT})^0{}_{Ai} \,,
 }
that are rotated into each other by a $\grU(1)_R$ symmetry.  On the complex combination $j_{\mu A}^\text{CFT} = j_{\mu A 1}^\text{CFT} + i  j_{\mu A 2}^\text{CFT}$, $\grU(1)_R$ acts as a phase rotation, $j_{\mu A}^\text{CFT} \to e^{i \theta} j_{\mu A}^\text{CFT}$, defined such that the complex supercurrent $j_{\mu A}^\text{CFT}$ and the corresponding supercharge $Q_A^\text{CFT}$ have R-charge $1$.  In this theory, $\grU(1)_R$ can be identified as the $\grU(1)$ symmetry that acts by a phase rotation on the complex combination $\widetilde{\psi}_1 + i \widetilde{\psi}_2$.  The complex ${\cal N}=(2, 2)$ supercurrent has the form \cite{Gopakumar:2012gd}:
\es{CFTSupercurrents}{
   j_{--}^\text{CFT} &=  \frac{1}{3 \sqrt 2} \tr \left ( \widetilde{\psi}_{1-} + i \widetilde{\psi}_{2-}\right )^3  \,,
       \qquad j_{+-}^\text{CFT} = 0 \,, \\
   j_{++}^\text{CFT} &= \frac{1}{3 \sqrt 2} \tr \left ( \widetilde{\psi}_{1+} + i \widetilde{\psi}_{2+}\right )^3 \,, 
       \qquad  j_{-+}^\text{CFT} = 0 \,,
 } 
where the normalization was chosen such that the anti-commutators of, say, $Q_{A1}$ with $j_{\mu A 1}$ produce the stress tensor \eqref{GotDeltaTAgain} also with a coefficient of $3N/(8 \pi)$.  From the $\grU(1)_R$ transformation of the supercurrent, we see that on the fermions the $\grU(1)_R$ must act as $\widetilde{\psi}_1 + i \widetilde{\psi}_2 \to e^{i \theta/ 3} (\widetilde{\psi}_1 + i \widetilde{\psi}_2)$, so the complex combination $\widetilde{\psi}_1 + i \widetilde{\psi}_2$ has R-charge $1/3$.

To form the total ${\cal N} = (1, 1)$ supercurrent, we can add one of the components of the ${\cal N} = (2, 2)$ supercurrent, for instance $j_{\mu A 1}$, to \eqref{MassiveSupercurrents}.  Then $j^{\rm tot}_{\mu A}= j_{\mu A} +  j^{\rm CFT}_{\mu A 1}$ has the explicit form
\es{ThreeadjSuper}{
   j_{--}^{\rm tot} &=  \frac{1}{3} \tr \psi_-^3 +  \tr ( \psi_- \widetilde{\psi}_{i-} \widetilde{\psi}_{i-} ) 
+ \frac{1}{3 \sqrt 2} \tr \widetilde{\psi}_{1-}^3 - \frac{1}{\sqrt 2} \tr \widetilde{\psi}_{1-} \widetilde{\psi}_{2-} \widetilde{\psi}_{2-}\,,
       \qquad j_{+-}^{\rm tot} = \alpha \tr \left( \psi_+ F \right) \,, \\
   j_{++}^{\rm tot} &= \frac{1}{3} \tr \psi_+^3 + \tr ( \psi_+ \widetilde{\psi}_{i+} \widetilde{\psi}_{i+} )
+ \frac{1}{3 \sqrt 2} \tr \widetilde{\psi}_{1+}^3 - \frac{1}{\sqrt 2} \tr \widetilde{\psi}_{1+} \widetilde{\psi}_{2+} \widetilde{\psi}_{2+} \,, 
       \qquad  j_{-+}^{\rm tot} = \alpha \tr \left( \psi_- F \right) \,.
 }
The stress-energy tensor $T_{--}$ calculated from anticommutator of $j_{--}$ given in (\ref{MassiveSupercurrents}) contains an extraneous 4-fermion term
$\sim \tr  \widetilde{\psi}_{1-} \widetilde{\psi}_{1-} \widetilde{\psi}_{2-} \widetilde{\psi}_{2-}$, but it is exactly canceled by the contribution from the CFT supercurrent anticommutator!
As a result, the stress energy tensor calculated from $j^{\rm tot}_{\mu A}$ given in (\ref{ThreeadjSuper}) does not contain any 4-fermion terms and has the canonical form. This demonstrates the ${\cal N}=(1, 1)$ supersymmetry of the full gauge theory, including both the massive and the CFT sector. 
Furthermore, the CFT has an enhanced ${\cal N}=(2, 2)$ supersymmetry \cite{Gopakumar:2012gd,Isachenkov:2014zua}.

As in Section~\ref{GENERALGAUGE}, this theory can be generalized to arbitrary gauge group $G$, with the modifications described there.  The infrared limit is described by the coset $SO(2\, \text{dim}\, G)_1 / G_{2 h^\vee}$, which again has ${\cal N} = (2, 2)$ supersymmetry, with complex supercurrent 
\es{CFTSupercurrents2}{
   j_{--}^\text{CFT} &=  \frac{i f^{abc}}{6 \sqrt 2} \left ( \widetilde{\psi}^a_{1-} + i \widetilde{\psi}^a_{2-}\right )
   \left ( \widetilde{\psi}^b_{1-} + i \widetilde{\psi}^b_{2-}\right )
   \left ( \widetilde{\psi}^c_{1-} + i \widetilde{\psi}^c_{2-}\right )  \,,
       \qquad j_{+-}^\text{CFT} = 0 \,, \\
   j_{++}^\text{CFT} &=  \frac{i f^{abc}}{6 \sqrt 2} \left ( \widetilde{\psi}_{1+}^a + i \widetilde{\psi}_{2+}^a\right ) 
      \left ( \widetilde{\psi}_{1+}^b + i \widetilde{\psi}_{2+}^b\right )
       \left ( \widetilde{\psi}_{1+}^c + i \widetilde{\psi}_{2+}^c\right ) \,, 
       \qquad  j_{-+}^\text{CFT} = 0 \,.
 } 
The canonical stress tensor of the full theory can be generated from $j^{\rm tot}_{\mu A}= j_{\mu A} +  j^{\rm CFT}_{\mu A 1}$, with $j_{\mu A}$ as in \eqref{LCsupercurrentsGen2}.

The construction presented in this section can be extended to
$\SU(N)$ gauge theory coupled to one massive adjoint and $N_f>2$ massless ones.  The low-energy limit is described by the coset CFT $\frac{\SO(N_f(N^2-1))_1}{\SU(N)_{N_f N}}$ of Virasoro central charge $\frac{N_f(N_f-1)(N^2-1)}{2(N_f+1)}$. By constructing the supercharge, we can show that that full gauge theory, including the massive sector, has
${\cal N}=(1,1)$ supersymmetry if the adjoint mass is taken to be $\sqrt\frac{(N_f+1) g^2 N}{2\pi}$. The details of this construction will be presented in a later publication.

\section*{Acknowledgments}

We are grateful to Ross Dempsey and Fedor Popov for very useful discussions. This work was supported in part by the Simons Foundation Grant No.~917464 (Simons Collaboration on Confinement and QCD Strings), the US Department of Energy under Award No.~DE-SC0007968, and the US National Science Foundation under Grants No.~PHY-2207584 and PHY-2209997.

\appendix

\section{Anomaly for composite operators}
\label{ANOMALYAPPENDIX}

We will present two derivations of \eqref{DerAB1} and \eqref{DerAB2}.  In the first derivation, presented in Section~\ref{ARBITRARYEPS}, we obtain directly the derivatives in \eqref{DerAB1} and \eqref{DerAB2}.  In the second derivation, presented in Section~\ref{EQUALTIME}, we work in equal-time quantization and with point-splitting in the space direction.  In addition to the derivatives \eqref{DerAB1} and \eqref{DerAB2} also obtain the necessary commutation relations that will be used for constructing the supercurrent multiplet.

\subsection{Anomaly for arbitrary $\epsilon^\mu$}
\label{ARBITRARYEPS}

Let us start with the gauge-invariant operator ${\cal X} = \tr ({\cal A} {\cal B})$.   We regularize this operator by point splitting, with a Wilson line inserted:
 \es{OabReg}{
  {\cal X}_\text{reg}(x) &= \frac 12 {\cal A}^a(x) U^{ac}(x, x + \epsilon) {\cal B}^c(x+\epsilon) \\
  &\approx \frac 12 {\cal A}^a(x) \left( \delta^{ac} - i \int_x^{x + \epsilon} dz^\nu\, A^b_\nu(z) (T^b)^{ac} \right) {\cal B}^c(x + \epsilon) \,,
 } 
where $(T^b)^{ac} = i f^{abc}$ are the generators in the adjoint representation.  Taking a derivative with respect to $x^\mu$, we find
 \es{DerOabReg}{
  \partial_\mu {\cal X}_\text{reg}(x) &\approx \frac 12 \partial_\mu {\cal A}^a(x) \left( \delta^{ac} + \int_x^{x + \epsilon} dz^\nu\, A^b_\nu(z) f^{abc} \right) {\cal B}^c(x + \epsilon) \\
  &{}+\frac 12 {\cal A}^a(x) \left( \delta^{ac} + \int_x^{x + \epsilon} dz^\nu\, A^b_\nu(z) f^{abc} \right) \partial_\mu {\cal B}^c(x + \epsilon) \\
  &{}+\frac 12 {\cal A}^a(x) f^{abc} \, \partial_\mu A_\nu^b(x) \epsilon^\nu \,   {\cal B}^c(x + \epsilon) \,.
 } 
In the first two lines, we can use the definitions of the covariant derivatives $D_\mu {\cal A}^a = \partial_\mu {\cal A}^a + f^{ade} A_\mu^d {\cal A}^e$ and $D_\mu {\cal B}^c = \partial_\mu {\cal B}^c + f^{cde} A_\mu^d {\cal B}^e$ to replace the regular derivatives in \eqref{DerOabReg} by covariant derivatives plus additional terms.   Up to quadratic terms in $\epsilon$, these additional terms combine with the last line of \eqref{DerOabReg} into a single term containing the gauge field strength.  The full expression is
 \es{DerOabRegAgain}{
  \partial_\mu {\cal X}_\text{reg}(x) &\approx \frac 12 ((D_\mu {\cal A}^a) {\cal B}^a)_\text{reg}
   + \frac 12 ({\cal A}^a D_\mu {\cal B}^a)_\text{reg} + \frac 12 {\cal A}^a(x) f^{abc} F^b_{\mu\nu}(x) \epsilon^\nu  {\cal B}^c(x + \epsilon) \,.
 } 

Let us now apply this equation to the case where the OPE between ${\cal A}$ and ${\cal B}$ is given by the first expression in \eqref{OPE2}.  Eq.~\eqref{DerOabRegAgain} implies
 \es{DerXCase1}{
  \partial_\mu {\cal X}(x) &= \tr  ((D_\mu {\cal A}) {\cal B}) +  \tr  ({\cal A} D_\mu {\cal B})
    - \frac 12 f^{acd}  f^{abc}  F^b_{\mu\nu}(x)  {\cal C}^d(x) \lim_{\epsilon \to 0}  \frac{\epsilon^\nu}{\epsilon^-}   \,.
 } 
With $ f^{acd}  f^{abc} = - N \delta^{bd}$, this can be written explicitly as
 \es{DerXCase1Explicit}{
  \partial_+ {\cal X}(x) &= \tr  ((D_+ {\cal A}) {\cal B}) +  \tr  ({\cal A} D_+ {\cal B})
    +N \tr (F {\cal C})  \,, \\
  \partial_- {\cal X}(x) &= \tr  ((D_- {\cal A}) {\cal B}) +  \tr  ({\cal A} D_- {\cal B})
    -N \tr (F {\cal C}) \lim_{\epsilon \to 0}  \frac{\epsilon^+}{\epsilon^-}   \,.  
 } 
If the OPE between ${\cal A}$ and ${\cal B}$ is given by the second expression in \eqref{OPE2}, we have
  \es{DerXCase2}{
  \partial_\mu {\cal X}(x) &= \tr  ((D_\mu {\cal A}) {\cal B}) +  \tr  ({\cal A} D_\mu {\cal B})
    + \frac 12 f^{acd}  f^{abc}  F^b_{\mu\nu}(x)  {\cal C}^d(x) \lim_{\epsilon \to 0}  \frac{\epsilon^\nu}{\epsilon^+}   \,,
 } 
which implies 
 \es{DerXCase2Explicit}{
  \partial_+ {\cal X}(x) &= \tr  ((D_+ {\cal A}) {\cal B}) +  \tr  ({\cal A} D_+ {\cal B})
    -N \tr (F {\cal C})  \lim_{\epsilon \to 0} \frac{\epsilon^-}{\epsilon^+}   \,, \\
  \partial_- {\cal X}(x) &= \tr  ((D_- {\cal A}) {\cal B}) +  \tr  ({\cal A} D_- {\cal B})
    + N \tr (F {\cal C})  \,.
 } 
The equations quoted in \eqref{DerAB2} in the main text are the first equation in \eqref{DerXCase1Explicit} and the second equation in \eqref{DerXCase2Explicit}.  These equations are independent of the direction of $\epsilon^\mu$.

Note that the second equation in \eqref{DerXCase1Explicit} and the first equation in \eqref{DerXCase2Explicit} depend on regularization.  In particular, if the point splitting is performed in the spatial direction, then we can use $\lim_{\epsilon \to 0}  \frac{\epsilon^\pm}{\epsilon^\mp} = -1$, while if we Lorentz average over the possible orientations of $\epsilon^\mu$, we can use \eqref{LIAvg} to conclude that $\lim_{\epsilon \to 0}  \frac{\epsilon^\pm}{\epsilon^\mp} = 0$.

A very similar analysis can be done for the derivatives of the operators ${\cal Y}$.  The result is that in the case where the OPE between ${\cal A}$ and ${\cal B}$ is given by the first expression in \eqref{OPE1}, we have
 \es{DerYCase1Explicit}{
  (D_+ {\cal Y})^a &= i f^{abc} (D_+ {\cal A})^b {\cal B}^c + i f^{abc} {\cal A}^b (D_+ {\cal B})^c - i  N  F^a {\cal C} \,, \\
  (D_- {\cal Y})^a &= i f^{abc} (D_- {\cal A})^b {\cal B}^c + i f^{abc} {\cal A}^b (D_- {\cal B})^c + i  N  F^a {\cal C} \lim_{\epsilon \to 0} \frac{\epsilon^+}{\epsilon^-} \,,
 }
while in the case where the OPE between ${\cal A}$ and ${\cal B}$ is given by the second expression in \eqref{OPE1}, we have
 \es{DerYCase2Explicit}{
  (D_+ {\cal Y})^a &=  i f^{abc} (D_- {\cal A})^b {\cal B}^c + i f^{abc} {\cal A}^b (D_- {\cal B})^c + i  N  F^a {\cal C} \lim_{\epsilon \to 0} \frac{\epsilon^-}{\epsilon^+} \,, \\
  (D_- {\cal Y})^a &=  i f^{abc} (D_- {\cal A})^b {\cal B}^c + i f^{abc} {\cal A}^b (D_- {\cal B})^c - i  N  F^a {\cal C}  \,.
 }
The equations quoted in \eqref{DerAB1} in the main text are the first equation in \eqref{DerYCase1Explicit} and the second equation in \eqref{DerYCase2Explicit}.  These equations are also independent of the direction of $\epsilon^\mu$.

\subsection{Anomaly using spatial point-splitting and equal-time quantization}
 \label{EQUALTIME}

In our second approach, we derive the anomaly equations \eqref{DerAB1} and \eqref{DerAB2} in two steps.  In the first step, we show that the commutator between the electric field operator $F^a(x)$ and  the composite operators  \eqref{ABReg} receives a contribution from the small Wilson line that was used for regularization.  In the second step, we derive \eqref{DerAB1} and \eqref{DerAB2} from the commutator of the composite operators \eqref{ABReg} with the momentum operators \eqref{GotPpm}.

It is perhaps most illuminating to work in canonical quantization in the gauge $A_0 = 0$.  In this gauge, the remaining component of the gauge field $A_1$ is canonically conjugate to the electric field, and its equal-time commutation relation with the field strength $F(x)$ reads
 \es{AFCommut}{
  [F^a(x), A_1^b(y)] = i g^2 \delta(x^1-y^1) \delta^{ab} \,, \qquad \text{for $x^0 = y^0$} \,.
 }

Let us now consider a composite operator of the form ${\cal O}^{ab} = {\cal A}^a {\cal B}^b$.  We regularize this operator by point splitting in the space direction, with $\epsilon^\mu = (0, \epsilon^1)$:
 \es{OabReg2}{
  {\cal O}^{ab}_\text{reg}(x) &= {\cal A}^a(x) U^{be}(x, x + \epsilon) {\cal B}^e(x+\epsilon)  \\
  &\approx {\cal A}^a(x) \left( \delta^{be} - i \int_x^{x + \epsilon} dz^\nu\, A^d_\nu(z) (T^d)^{be} \right) {\cal B}^e(x + \epsilon) \,,
 } 
with $(T^d)^{be} = i f^{bde}$ as before.

Let us now compute the commutator of $F^c(x)$ and ${\cal O}^{ab}(y)$ at equal times $x^0 = y^0$.  We have
  \es{CommutOabReg}{
  [F^c(x), {\cal O}^{ab}_\text{reg}(y)] &\approx [F^c(x), {\cal A}^a(y)] U^{be}(y, y + \epsilon){\cal B}^e(y+\epsilon)  \\
   &{}+{\cal A}^a(y) U^{be}(y, y + \epsilon) [F^c(x), {\cal B}^e(y+\epsilon)] \\
   &{}+ i g^2 {\cal A}^a(y)  f^{bce} {\cal B}^e(y + \epsilon) \times
    \begin{cases}
      1 & \text{if $y^1 < x^1 < y^1 + \epsilon^1$} \\
      0 & \text{otherwise} 
    \end{cases} \,.
 } 

As we take $\epsilon^1 \to 0$, we can write this expression as
 \es{CommutOab}{
  [F^c(x), {\cal O}^{ab}(y)] &= [F^c(x), {\cal A}^a(y)] {\cal B}^b(y) + {\cal A}^a(y)[F^c(x), {\cal B}^b(y)] \\
  &{}+ i g^2 f^{bce} \delta(x^1-y^1) \lim_{\epsilon^1 \to 0} \left( \epsilon^1 {\cal A}^a(y) {\cal B}^e(y + \epsilon) \right) \,,
 }
where the first term is the composite of $[F^c(x), {\cal A}^a(y)]$ with ${\cal B}^b(y)$, and in the second line we have the composite of ${\cal A}^a(y)$ with $[F^c(x), {\cal B}^b(y)]$, both composites being regularized by gauge-invariant point splitting.

We can apply this equation to the two composite operators defined in \eqref{ABReg}.  The first operator is ${\cal X} = \frac 12 {\cal O}^{aa}$, and in this case the OPE \eqref{OPE2} implies
 \es{OPE2Limit}{
  \lim_{\epsilon^1 \to 0} \left( \epsilon^1 {\cal A}^a(y) {\cal B}^e(y + \epsilon) \right)  = \sqrt{2} f^{aed} {\cal C}^d(y)   \,.
 }
This implies the equal-time commutator
 \es{CommutO1}{
  [F^c(x), {\cal X} (y)] &= \frac 12 [F^c(x), {\cal A}^a(y)] {\cal B}^a(y) + \frac 12 {\cal A}^a(y)[F^c(x), {\cal B}^a(y)] \\
  &{}- \frac{i}{\sqrt{2}} g^2 N   \delta(x^1-y^1)  {\cal C}^c(y) \,,
 }
where we used $f^{ace} f^{aed} = - N \delta^{cd}$.

Similarly, the second composite operator in \eqref{ABReg} has color components $ {\cal Y}^d = i f^{dab} {\cal O}^{ab}$.  In this case the OPE \eqref{OPE1} implies
 \es{OPE1Limit}{
  \lim_{\epsilon^1 \to 0} \left( \epsilon^1 {\cal A}^a(y) {\cal B}^e(y + \epsilon) \right)  = \sqrt{2} \delta^{ae} {\cal C}(y)   \,.
 }
The equal-time commutator with $F^c$ is then
 \es{CommutO2}{
  [F^c(x), {\cal Y}^d (y)] &= i f^{dab} [F^c(x), {\cal A}^a(y)] {\cal B}^b(y) + i f^{dab} {\cal A}^a(y)[F^c(x), {\cal B}^b(y)] \\
  &{}-   \sqrt{2}  \delta^{cd} g^2 N  \delta(x^1-y^1) {\cal C}(y)   \,.
 }

Since both $P_+$ and $P_-$ contain a term $\int dx^1\, \frac{1}{\sqrt{2} g^2} \tr F^2(x)$ (see \eqref{GotPpm}), the commutators of $P_\pm$ with ${\cal X}$ and ${\cal Y}$ will have additional terms coming from the last terms in \eqref{CommutO1} and \eqref{CommutO2}:
 \es{PpmCommut}{
  [P_\pm, {\cal X}] &= \tr ([P_\pm, {\cal A}] {\cal B}) + \tr ({\cal A} [P_\pm, {\cal B}]) - i N \tr (F {\cal C}) \,, \\
  [P_\pm, {\cal Y}^a] &= i f^{abc} \left([P_\pm, {\cal A}^b] {\cal B}^c + {\cal A}^b [P_\pm, {\cal B}^c] \right) - N  F^a {\cal C} \,. 
 }
Since $D_\mu {\cal O} = i [P_\mu, {\cal O}]$ for any operator ${\cal O}$, the relations \eqref{PpmCommut} immediately imply \eqref{DerAB1} and \eqref{DerAB2}.

\subsection{Commutator between $F^a$ and $\tr \psi_-^3$ and $\tr \psi_+^3$}

Let us now use the formalism introduced above to derive the commutation relations \eqref{FAdditional} mentioned in the main text.  For the first relation, we have
 \es{Ftrpsi3}{
  \left[F^a(t, x), \tr (\psi_- J_-)(t, y)\right] =  \frac 12 \psi_-^b(t, y)[F^a(t, x), J_-^b(t, y)] + \frac{i}{2\sqrt{2}\pi} g^2 N   \delta(x-y)  \psi_-^a(t, y) \,,
 }
which follows from \eqref{CommutO1} with ${\cal A} = \psi_-$, ${\cal B} = J_-$, and ${\cal C} = - \frac{1}{2 \pi} \psi_-$ (see also the text around \eqref{psiJmOPE}).  From \eqref{CommutO2} with ${\cal A} = {\cal B} = \psi_-$ and ${\cal C} = - \frac{i}{2 \pi}$ (see the text before \eqref{DJp}), we have
 \es{FJ}{
  [F^a(t, x), J_-^b(t, y)] =  \frac{i}{2\sqrt{2}\pi} \delta^{ab} g^2 N \delta(x-y) \,.
 }
Plugging \eqref{FJ} into \eqref{Ftrpsi3}, we find
 \es{Ftrpsi3Final}{
  \left[F^a(t, x), \frac 13 \tr (\psi_-^3)(t, y)\right] =   \frac{ig^2 N}{4\sqrt{2}\pi}   \delta(x-y)  \psi_-^a(t, y) \,.
 }
The second relation in \eqref{FAdditional} is derived in a similar manner.

\section{Relevant commutators and anti-commutators}
\label{COMMUTATORS}

The anti-commutators in \eqref{CanRels} and the commutators in \eqref{FAdditional} imply the following equal time commutation relations
 \es{RelevantComm}{
    \left[ \frac{1}3 \tr \left( \psi_-^3  \right)(t, x),  \tr (F^2)(t, y) \right]
   &=  -\frac{ i g^2 N}{2 \pi \sqrt{2} } \tr (\psi_- F)(t, x) \delta(x-y) \,, \\
    \left[ \frac{1}3 \tr \left( \psi_+^3  \right)(t, x),  \tr (F^2)(t, y) \right]
   &=  \frac{ i g^2 N}{2 \pi \sqrt{2} } \tr (\psi_+ F)(t, x) \delta(x-y)  \,,
 }
as well as the anti-commutation relations
 \es{RelevantAnticomm}{
  \left\{ \frac{1}3 \tr \left( \psi_-^3 \right)(t, x), \frac{1}3 \tr \left( \psi_-^3  \right)(t, y) \right\}
    &= -\frac{ N(N^2 - 1)}{48 \sqrt{2} \pi^2} \delta''(x-y) \mathds{1} \\
    &{}-\frac{i N}{4\pi} \tr \left( \psi_- D_1 \psi_- \right )(t, x) \delta(x-y) \,, \\
  \left\{ \frac{1}3 \tr \left( \psi_-^3  \right)(t, x),  \tr (F \psi_+)(t, y) \right\}
   &=  \frac{ i g^2 N}{4 \pi \sqrt{2} } \tr (\psi_+ \psi_-)(t, x) \delta(x-y) \,, \\
  \left\{ \frac{1}3 \tr \left( \psi_-^3  \right)(t, x),  \tr (F \psi_-)(t, y) \right\}
   &= \frac{1}{\sqrt{2}} \tr (F \psi_- \psi_-)(t, x) \delta(x-y) \,, \\
  \left\{ \frac{1}3 \tr \left( \psi_+^3 \right)(t, x), \frac{1}3 \tr \left( \psi_+^3  \right)(t, y) \right\} 
   &= -\frac{ N(N^2 - 1)}{48 \sqrt{2} \pi^2} \delta''(x-y) \mathds{1} \\
   &{}+\frac{i N}{4\pi} \tr \left( \psi_+ D_1 \psi_+ \right )(t, x) \delta(x-y) \,, \\
    \left\{ \frac{1}3 \tr \left( \psi_+^3  \right)(t, x),  \tr (F \psi_+)(t, y) \right\}
   &=  \frac{1}{\sqrt{2}} \tr (F \psi_+ \psi_+)(t, x) \delta(x-y) \,, \\
  \left\{ \frac{1}3 \tr \left( \psi_+^3  \right)(t, x),  \tr (F \psi_-)(t, y) \right\}
   &=  \frac{ i g^2 N}{4 \pi \sqrt{2} } \tr (\psi_+ \psi_-)(t, x) \delta(x-y)  \,, \\
  \{ \tr (F \psi_-)(t, x), \tr (F \psi_-)(t, y) \} &=  \frac{1}{\sqrt{2}} \tr (F^2)(t, x) \delta (x-y) \,, \\
  \{ \tr (F \psi_+)(t, x), \tr (F \psi_+)(t, y) \} &=  \frac{1}{\sqrt{2}} \tr (F^2)(t, x) \delta (x-y) \,.
 }
All these equations are straightforward to derive from \eqref{CanRels} and \eqref{FAdditional} using standard commutation identities, except for the first and the fourth equations.  Let us explain the derivation of the first equation (the derivation of the fourth equation proceeds analogously), starting from the theory of $N^2 -1 $ free massless Majorana fermions, where we do not have to worry about the Wilson lines that are needed in the gauge-invariant point-splitting regularization.  Let us start with the OPE
 \es{OPEpsi3psi3}{
  \frac{1}3 \tr \left( \psi_-^3 \right)(x)\,  \frac{1}3 \tr \left( \psi_-^3  \right)(0)
   &= \frac{i N(N^2 - 1)}{24} \frac{1}{(2 \pi x^-)^3} \mathds{1} \\
    &{}+ \frac{N}{8}  \frac{1}{(2 \pi x^-)^2} : \psi_-^a(0) \psi_-^a(x): \\
    &{}+ \frac{i}{16} f^{abc} f^{ade} \frac{1}{2 \pi x^-}  :\psi_-^b(x) \psi_-^c(x) \psi_-^d(0) \psi_-^e(0): \\
    &{}+ \text{(regular)} \,,
 }
where the normal ordering symbols mean that we do not perform self-contractions.   The equal-time anti-commutator of fermionic operators $X$ and $Y$ can be obtained from the OPE via
 \es{AntiComm}{
  \left\{ X(0, x), Y(0) \right\}
   = \lim_{\substack{\epsilon \to 0 \\ \epsilon> 0 }}
    \left[ X(- i \epsilon, x) Y(0) - X( i \epsilon, x) Y(0) \right] \,,
 }
where $\epsilon > 0$. Applying this rule to \eqref{OPEpsi3psi3} we find
 \es{psi3psi3Anti}{
  \left\{ \frac{1}3 \tr \left( \psi_-^3 \right)(0, x), \frac{1}3 \tr \left( \psi_-^3  \right)(0) \right\}
   &= \lim_{\substack{\epsilon \to 0 \\ \epsilon> 0 }}
    \Biggl[ -\frac{i N(N^2 - 1)}{6 \sqrt{2} (2 \pi)^3}  \left[  \frac{1}{(x + i \epsilon)^3} - \frac{1}{(x - i \epsilon)^3} \right]  \mathds{1} \\
   &{}+\frac{N}{4 (2 \pi)^2} : \psi_-^a\partial_1 \psi_-^a(0): \left[ \frac{x}{(x + i \epsilon)^2} 
    - \frac{x}{(x - i \epsilon)^2} \right] \\
   &{}+\frac{N}{4 (2 \pi)^2} : \psi_-^a\partial_0 \psi_-^a(0):  \left[ \frac{-i \epsilon}{(x + i \epsilon)^2} 
    - \frac{i \epsilon }{(x - i \epsilon)^2} \right]   \\
   &{}-\frac{i f^{abc} f^{ade}}{16 \sqrt{2} \pi }   :\psi_-^b \psi_-^c \psi_-^d \psi_-^e(0): 
    \left[ \frac{1}{x + i \epsilon} 
    - \frac{1}{x - i \epsilon}  \right] \Biggr] \,,
   }
where we kept only the terms that are potentially singular.  Taking the $\epsilon \to 0$ limit and noticing that $f^{abc} f^{ade} :\psi_-^b \psi_-^c \psi_-^d \psi_-^e(0): $ vanishes due to the Jacobi identity, we find
 \es{psi3psi3Anti2}{
   \left\{ \frac{1}3 \tr \left( \psi_-^3 \right)(0, x), \frac{1}3 \tr \left( \psi_-^3  \right)(0) \right\}
    = -\frac{ N(N^2 - 1)}{48 \sqrt{2} \pi^2} \delta''(x) \mathds{1} 
     -\frac{i N}{8 \pi} \delta(x)  : \psi_-^a\partial_1 \psi_-^a(0): \,.
 }
The first term is related by supersymmetry to a corresponding term that appears in the commutator of two stress tensors, and whose coefficient is proportional to the ultraviolet Virasoro central charge.  Because it is a total derivative, the first term in \eqref{psi3psi3Anti2} does not contribute to the anti-commutator of the supercharges with the supercurrent.  The expression \eqref{psi3psi3Anti2} is precisely what appears in the first line of \eqref{RelevantAnticomm} after appropriately turning regular derivatives into covariant derivatives.  Note that since the leading term in the OPE of the fermionic operators should be completed into a gauge-covariant expression (see the discussion after \eqref{OPE1}), the final expression for the anti-commutator of the gauge-invariant operators in the first line of \eqref{RelevantAnticomm} should be gauge-invariant.  A similar comment applies to all expressions in \eqref{RelevantComm} and \eqref{RelevantAnticomm}, namely all operators on the right-hand sides are gauge-invariant, and the composite operators are regularized by gauge-invariant point splitting in the space direction.

The commutation relations \eqref{RelevantComm} imply 
 \es{QO}{
  [Q_-, Z ] =  -\frac{ i \alpha N}{8 \pi } \tr (\psi_- F) =  -\frac{i N}{8 \pi } j_{-+} \,, \qquad
   [Q_+, Z ] =  \frac{i \alpha N}{8 \pi } \tr (\psi_+ F) = \frac{i N}{8 \pi } j_{+-}  \,,
 }
so indeed, by acting with $Q_\pm$ on $Z$ we obtain components of the supercurrent.  Acting on the supercurrent components with $Q_-$ gives 
 \es{Qplusj}{
  \{ Q_-, j_{--}\} &= \frac{1}{\sqrt{2}} \left[ -\frac{i N}{4\pi} \tr \left( \psi_- D_1 \psi_- \right )
   -  \frac{ i  N m}{8 \pi  } \tr (\psi_- \psi_+) \right]  = \frac{i N}{8 \pi} \tr \left( \psi_- D_- \psi_- \right ) \,, \\
  \{ Q_-, j_{-+}\} &= \frac{\alpha}{2} \tr (F \psi_- \psi_-) 
   = \frac{\alpha}{4 g^2} \partial_- \tr (F^2)\,, \\
  \{ Q_-, j_{+-} \} &=  
    \frac{N}{8 \pi}  \left[ \frac{1}{g^2} \tr (F^2) +  \frac{ i m}{\sqrt{2} } \tr (\psi_+ \psi_-) \right] \,, \\
  \{ Q_-, j_{++} \} &=   \frac{\alpha}{2} \tr (F \psi_+ \psi_+) 
   = -\frac{\alpha}{4 g^2} \partial_+ \tr (F^2)\,,
 }
where in the first line we used the equation of motion to write $D_1 \psi_- = \frac{D_+ \psi_- - D_- \psi_-}{\sqrt{2}} = - \frac{m}{2} \psi_+ - \frac{1}{\sqrt{2}} D_- \psi_-$ as well as $\tr (F \psi_- \psi_-) = \frac{1}{2g^2} \partial_- \tr (F^2)$ and $\tr (F \psi_+ \psi_+) = -\frac{1}{2g^2} \partial_+ \tr (F^2)$.  Similarly, acting on the supercurrent with $Q_+$ gives
 \es{Qminusj}{
  \{Q_+, j_{++} \} &= \frac{1}{\sqrt{2}} \left[  \frac{i N}{4\pi} \tr \left( \psi_+ D_1 \psi_+ \right )
    +  \frac{ i  N m}{8 \pi  } \tr (\psi_+ \psi_-) \right] = \frac{i N}{8 \pi} \tr \left( \psi_+ D_+ \psi_+ \right ) \,, \\
  \{ Q_+, j_{+-}\} &= \frac{\alpha}{2} \tr (F \psi_+ \psi_+) = -\frac{\alpha}{4 g^2} \partial_+ \tr (F^2) \,, \\ 
  \{ Q_+, j_{-+} \} &=  
    \frac{N}{8 \pi}  \left[ \frac{1}{g^2} \tr (F^2) +  \frac{ i m}{\sqrt{2} } \tr (\psi_+ \psi_-) \right] \,, \\ 
   \{ Q_+, j_{--} \} &=   \frac{\alpha}{2} \tr (F \psi_- \psi_-) 
    = \frac{\alpha}{4 g^2} \partial_- \tr (F^2) \,,
 }
where in the first line we wrote $D_1 \psi_+ = \frac{D_+ \psi_+ - D_- \psi_+}{\sqrt{2}} =  -\frac{m}{2} \psi_- + \frac{1}{\sqrt{2}} D_+ \psi_+$.

 Comparing with the formulas for the stress tensor \eqref{TComponents} and for the operator $Z$ in \eqref{ZDef}, we reproduce \eqref{QjExplicit}.

In the case of additional massless fermions, we will also need the following commutators
 \es{RelevantCommGen}{
    \left[ \tr \left( \psi_- \widetilde{J}_-  \right)(t, x),  \tr (F^2)(t, y) \right]
   &=  -\frac{ i g^2 k}{2 \pi \sqrt{2} } \tr (\psi_- F)(t, x) \delta(x-y) \,, \\
    \left[ \tr \left( \psi_+ \widetilde{J}_+  \right)(t, x),  \tr (F^2)(t, y) \right]
   &=  \frac{ i g^2 k}{2 \pi \sqrt{2} } \tr (\psi_+ F)(t, x) \delta(x-y)  \,,
 }
which follow as an immediate application of \eqref{FAdditional2}, and the following anti-commutators
 \es{RelevantAnticommGen}{
  \left\{ \tr \left( \psi_- \widetilde{J}_- \right)(t, x),  \tr \left( \psi_- \widetilde{J}_-  \right)(t, y) \right\}
   &= -\frac{ k(N^2 - 1)}{16 \sqrt{2} \pi^2} \delta''(x-y) \mathds{1} \\
   &{}+\frac{1}{\sqrt{2}} \tr (\widetilde{J}_- \widetilde{J}_-)(t, x) \delta(x-y)\,, \\
  \left\{ \tr \left( \psi_- \widetilde{J}_-  \right)(t, x),  \tr (F \psi_+)(t, y) \right\}
   &=  \frac{ i g^2 k}{4 \pi \sqrt{2} } \tr (\psi_+ \psi_-)(t, x) \delta(x-y) \,, \\
  \left\{ \tr \left( \psi_- \widetilde{J}_-  \right)(t, x),  \tr (F \psi_-)(t, y) \right\}
   &= \frac{1}{\sqrt{2}} \tr (F \widetilde{J}_-)(t, x) \delta(x-y) \,, \\
  \left\{ \tr \left( \psi_+ \widetilde{J}_+ \right)(t, x), \tr \left( \psi_+ \widetilde{J}_+  \right)(t, y) \right\} 
   &= -\frac{ k(N^2 - 1)}{16 \sqrt{2} \pi^2} \delta''(x-y) \mathds{1}\\
   &{}+\frac{1}{\sqrt{2}} \tr (\widetilde{J}_+ \widetilde{J}_+)(t, x) \delta(x-y) \,, \\
    \left\{  \tr \left( \psi_+ \widetilde{J}_+ \right)(t, x),  \tr (F \psi_+)(t, y) \right\}
   &=  \frac{1}{\sqrt{2}} \tr (F \widetilde{J}_+)(t, x) \delta(x-y) \,, \\
  \left\{  \tr \left( \psi_+ \widetilde{J}_+  \right)(t, x),  \tr (F \psi_-)(t, y) \right\}
   &=  \frac{ i g^2 k}{4 \pi \sqrt{2} } \tr (\psi_+ \psi_-)(t, x) \delta(x-y)  \,,
 }
which follow from \eqref{CanRels}, \eqref{FAdditional}, and \eqref{FAdditional2}.

\bibliographystyle{ssg}
\bibliography{Supercurrent}

\end{document}